# Integration of Liquid Thermoelectrochemical Conversion into Forced Convection Cooling

Yutaka Ikeda, Kazuki Fukui and Yoichi Murakami*

School of Engineering, Tokyo Institute of Technology, 2-12-1-I1-15 Ookayama, Meguro-ku, Tokyo 152-8552, Japan

*Corresponding Author: Yoichi Murakami, E-mail: murakami.y.af@m.titech.ac.jp

**Abstract**

Forced convection cooling is important in numerous technologies ranging from microprocessors in data centers to turbines and engines; active cooling is essential in these situations. However, active transfer of heat or thermal energy under a large temperature difference promptly destroys the *exergy*, which is the free-energy component of thermal energy, and this issue has remained unaddressed. Herein, we describe a thermoelectrochemical conversion to partially recover presently lost exergy in forced convection cooling. We design a test cell in which an electrolyte liquid is forced through a channel formed between two parallel electrodes and the hot-side electrode simulates an object to be cooled. Our investigations show that the narrower interelectrode channels afford higher cooling and power generation performances. The mass transfer resistance is the most dominant type of resistance for all the conditions tested and the charge transfer kinetics is likely to be controlled by viscosity. The dependence of the generated power on the flow rate is caused by the change in the diffusion coefficient of redox species with temperature. As an evaluation measure for such forced-flow thermocells, the gain (Λ)—defined as the ratio of the generated power to the hydrodynamic pumping work required to force the liquid through the cell—is introduced. Λ is above unity in a certain flow rate region. This demonstrates that such a system can generate more electric power than the pump work required to drive the liquid through the cell, suggesting its potential to partly recover presently lost exergy of thermal energy as electricity.

**Broader Context** (maximum 200 words)

In our present civilization, forced convection cooling is used in wide-ranging situations from cooling of microprocessors in data centers to that of heat engines including turbines and automobile engines. Active cooling is essential in such situations to avoid thermal failure (for microprocessors) and maximize fuel-to-work conversion efficiencies (heat engines). Here, "active cooling" means the prompt removal of a large quantity of thermal energy in the heat source by a working fluid under a large temperature difference. However, this causes rapid destruction of the *exergy*, which is the free-energy component of the thermal energy. This issue has remained unaddressed despite the widespread use of forced convection cooling. In this study, to partially recover presently lost exergy in such situations, we integrate thermoelectrochemical conversion, which has been mostly studied for stationary conditions, into forced convection cooling. Through experimental and numerical investigations of a test cell in which the hot-side electrode simulates an object to be cooled, several fundamental properties of such a forced-flow cell are obtained. Our results indicate that such a forced-flow thermocell can generate a larger electric power than the hydrodynamic pumping work required to force the liquid through the cell unit, justifying the concept of this kind of thermocell.

# 1 Introduction

Forced convection cooling is ubiquitous and fundamental to sustain our current civilization. For example, all heat engines including power turbines and automobile engines require forced convection cooling to maximize their fuel-to-work conversion efficiencies. This is because their efficiencies are thermodynamically limited by the Carnot efficiency $\eta_\text{C} = 1 - (T_\text{heat,out}/T_\text{heat,in})$,[1] where $T_\text{heat,in}$ and $T_\text{heat,out}$ are the temperatures of the heat source and release side, respectively, and thus $T_\text{heat,out}$ should be minimized to realize higher efficiencies. Microprocessors in data centers are often operated under forced convection cooling using a liquid coolant[2,3] to prevent their thermal failure. This is because the rate of failure generally obeys the Arrhenius function[4,5] and thus increases exponentially with temperature. The recent upsurge in the information and communications sector is increasing the number of data centers worldwide and their electricity consumption has reached 1%–2% of global electricity generation;[2,3] this percentage is predicted to increase further in the near future.[2] Forced convection cooling is effective for meeting cooling requirements but necessitates a pumping power to send the coolant through the heat dissipation section.

The aforementioned situations may be regarded as a loss of a huge amount of thermal energy. However, what is lost in such situations is not the thermal energy itself but *the exergy of thermal energy*,[6] which is the amount of free energy in the thermal energy.[1,6] Specifically, when a heat $\dot{Q}$ [J/s] is transferred from a heat-releasing object at $T_\text{surf}$ [K] toward the working fluid at a mean temperature of $T_\text{liq}$ [K], the rate of the loss of exergy $E_\text{X,loss}$ [J/s] is[1,6]

$$E_\text{X,loss} = \dot{Q} T_\text{env} \left( \frac{1}{T_\text{liq}} - \frac{1}{T_\text{surf}} \right) = \frac{T_\text{env}}{T_\text{liq} T_\text{surf}} \dot{Q} \Delta T', \qquad (1)$$

where $T_\text{env}$ is the temperature of the surrounding environment (~300 K) and $\Delta T' = T_\text{surf} - T_\text{liq}$. Equation (1) means that as $\dot{Q}$ and $\Delta T'$ become larger, $E_\text{X,loss}$ increases. Therefore, active cooling is physically equal to a rapid loss of a large amount of exergy. However, this aspect of forced convection cooling has remained unaddressed despite the widespread use of this technology.

Since the last century, liquid-based thermoelectric conversion, termed thermogalvanic[7–11] or thermoelectrochemical conversion,[12–14] has been explored for recovering waste thermal energy as electrical energy.[7-36] This conversion technology uses an electric potential difference ($\Delta E$) caused by redox reactions on two electrodes held at different temperatures. In liquid-based thermoelectric conversion, the temperature coefficient of $\Delta E$, or Seebeck coefficient $\partial E/\partial T$, is related to the reaction entropy $\Delta S_{rx}$ by[7,10,14]

$$\frac{\partial E}{\partial T} \approx \left|\frac{\Delta S_{rx}}{nF}\right|, \qquad (2)$$

where $n$ is the number of electrons involved in the reaction and $F$ is the Faraday constant. To date, numerous related studies using various kinds of solvents including water,[7,8,12,15,19-21,28-32,34] organic solvents,[9,18] and ionic liquids[13,16,17,22-27,33] have been reported. In particular, the use of ionic liquids, which are room-temperature molten salts with negligible vapor pressure and flammability,[37-40] has shown promise because ionic liquids can be applied to a broad range of heat sources above 100 °C. The use of ionic liquids in liquid-based thermoelectric conversion has been pioneered by the research groups headed by Pringle and MacFarlane,[13,22,23,25,26] Katayama,[16,17,24,33] and Cola.[27] New electrode materials such as carbon nanotubes,[12,19,20,28,30-32,36] reduced graphene oxide,[21] and poly(3,4-ethylenedioxythiophene) on stainless steel[25] have also been explored in liquid-based thermoelectric conversion. To date, most studies have been conducted using stationary thermocells in which the electrolyte liquid is encased in a closed vessel.[7-34] For example, certain stationary thermocells have been proposed as flexible alternatives[28] to solid-state thermoelectric materials.

In this report, to recover presently lost exergies of thermal energy in broad cooling situations, we describe an integration of liquid-based thermoelectrochemical conversion into forced convection cooling based on findings from both experiments and simulations. Since our initial demonstration of such an integration using our prototypical forced-flow thermocell,[41] we have repeated experiments to unambiguously confirm the data reproducibility and obtain reasonable interpretations that can consistently explain the results with the aid of our simulations. It should be noted that Cola et al.[35,36] have also proposed a similar idea and reported the results obtained using their flow-type thermocells but the concepts in our present work were obtained independently[41] from theirs. In their initial report,[35] they developed a cell with a serpentine channel into which 44 electrode rods were inserted from the top cover plate along the direction of the liquid flow. The aim of this geometry was to lower the mass transfer resistance by setting the direction of the flow and that of the array of rods parallel with each other. They demonstrated a harvesting of 2 μW between the first and second electrode rods from the inlet and the estimated total power was up to 88 μW.[35] In their subsequent report,[36] the same group used sheets of multi-walled carbon nanotubes as electrodes and attached the sheets on the top and bottom walls of the cell to obtain a large temperature gap between them; this setup successfully generated a power of 0.36 W/m$^2$. However, the cooling properties and electrochemical cell resistances were not presented in that report.

To bring the concept of integration of thermoelectrochemical conversion and forced convection cooling to practical uses, it is essential to elucidate their heat transfer, hydrodynamic, and electrochemical properties and understand their interdependence, which have yet to be elucidated. This report presents these properties, which are revealed through the investigation of our forced-flow thermocell equipped with different electrodes. The ionic liquid [C$_2$mim][NTf$_2$] (C$_2$mim = 1-ethyl-3-methylimidazolium, NTf$_2$ = bis(trifluoromethylsulfonyl)amide; Fig. 1a) is used as the solvent and Co$^{II/III}$(bpy)$_3$(NTf$_2$)$_{2/3}$ (bpy = 2,2′-bipyridine, Fig. 1b) as the redox couple.[22] In this report, the experimental results for cooling and power generation properties are presented first along with discussions supported by electrochemical experiments and numerical simulations. Then, we introduce and use dimensionless measures that are considered to be appropriate for evaluating such forced-flow thermocells whose principle function is forced convection cooling.

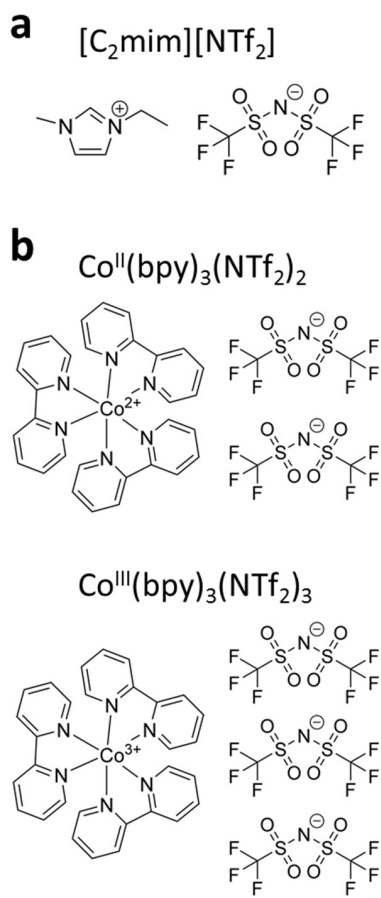

**Fig. 1** Molecular structures of the (a) ionic liquid [$C_2$mim][$NTf_2$] and (b) redox couple $Co^{II}(bpy)_3(NTf_2)_2$ and $Co^{III}(bpy)_3(NTf_2)_3$ used in this study.

## 2 Test Cell

In the cell we have designed and fabricated (Fig. 2), an electrolyte liquid is forced as a coolant through a channel formed between two parallel electrodes. The hot-side electrode (cathode)

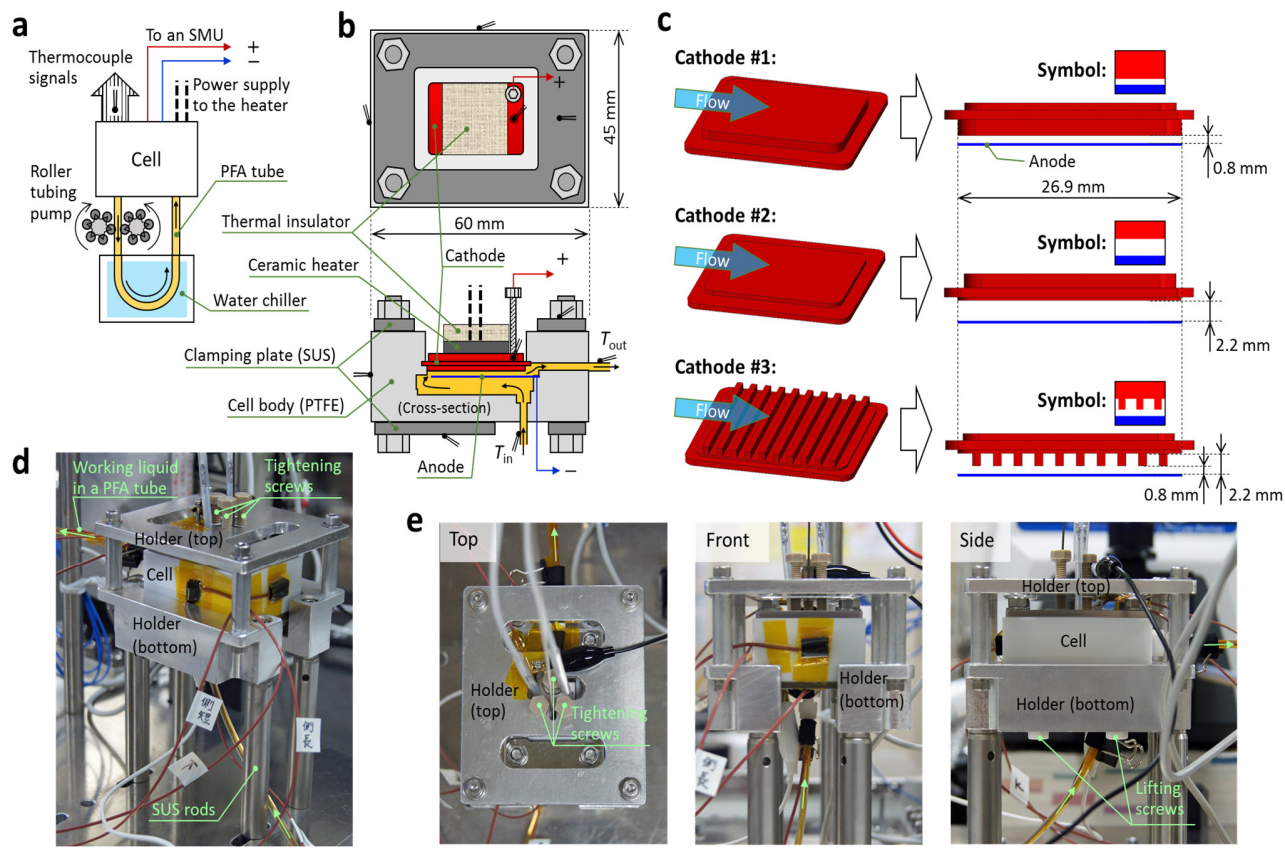

**Fig. 2** (a) Schematic of the experimental setup. (b) Top (upper) and cross-sectional (lower) illustrations of the cell drawn at scaled dimensions. (c) CAD graphics of cathode #1−3 (hot-side electrodes, red) and the interelectrode channels formed between the cathode and anode (cold-side electrode, blue). (d) A photograph of the cell in an aluminum holder mounted on four 0.5″-diameter SUS rods. (e) Photographs of the top (left), front (middle), and side (right) of the setup. The photographs in (d) and (e) were taken during an experiment. Light-green arrows indicate the flow direction and the brown wires are teflon-coated thermocouple wires.

simulates an object to be cooled. A 0.06 M solution of $Co^{II/III}(bpy)_3(NTf_2)_{2/3}$ in $[C_2mim][NTf_2]$ was used as the electrolyte and is termed the "working liquid" hereafter. Figure 2a schematically illustrates the experimental setup. The liquid was circulated using a roller tubing pump (Masterflex, Cole-Parmer; see Section 5) through a closed loop of tubes and tubing junctions made of perfluoroalkoxy alkane (PFA) and polytetrafluoroethylene (PTFE), respectively. The flow rate of the liquid ($G$) was set between ca. 0.1 and 0.5 mL/s. The water chiller was adjusted so that the liquid temperature at the entrance of the cell ($T_{in}$) was fixed at 25±2 °C. The cell had output leads
7

that were used for current–voltage (*I-V*) curve measurements with a source measure unit (SMU) and alternating current (AC) impedance measurements (see Section 5 for details).

Figure 2b illustrates top and cross-sectional views of the cell at scaled dimensions. The horizontal dimensions were 60 × 45 mm. The cell consisted of an anode (cold side, Pt plate), cathode (hot side, Pt-coated Ni), and ceramic heater to adjust the temperature of the cathode. Full details are provided in Section 5. The cell was constructed of three PTFE blocks and integrated using O-rings (not drawn) and stainless-steel (SUS) clamping plates tightened by bolts. The surface temperatures were recorded by the thermocouples and used to check the consistency with the results of our simulations (Section 5). As depicted in Fig. 2b, after the liquid enters the cell, it passes through the space beneath the anode and then enters an interelectrode channel between the anode and cathode (see Fig. S1 in ESI† for a magnified graphic of this cross section). Because of this flow, a steady-state temperature difference was established between the electrodes. Our simulation confirmed that most heat transfer to the working liquid occurred on the cathode plane during the passage of the liquid through the channel (Fig. S1c, ESI†).

We designed three cathodes denoted as #1−3 (Fig. 2c). Cathode #1 and 2 had flat surfaces with a liquid contact area of 5.9 cm$^2$ (cathode surfaces had lengths of 26.9 and 22.1 mm parallel and perpendicular to the flow, respectively). Cathode #3 had an extended surface with ten fins (width: 1 mm, pitch: 2.5 mm, height: 1.4 mm) and liquid contact area of 12.1 cm$^2$. The use of these cathodes resulted in different interelectrode channel geometries (Fig. 2c). We chose narrow channel spacings (0.8–2.2 mm) to increase the heat transfer coefficient (Section 2 of ESI† for the theoretical basis), which is the strategy used for microchannel heat sinks.[42-44] For all experiments in this report, the Reynolds number (*Re*) in the interelectrode channel was below 3 and thus the flow was highly laminar; i.e., turbulence was negligible.

Figures 2d and e show photographs of the cell in an aluminum holder. In the holder, the cell was lifted in the air by a few millimeters using four PTFE lifting screws, as can be seen in the side view in Fig. 2e (see Section 5 for details). With this cell holder, the physical and thermal contact between the cell and holder was minimized and the experimental reproducibility was enhanced. The ambient air temperature near the setup was monitored and controlled at 23±2 °C during all experiments.

In experiments presented below, the cathode temperature ($T_{cathode}$) was set between 70 and 170 °C with $T_{in}$ = 25±2 °C. We confirmed that the temperature non-uniformity on the cathode surface was minor, as assessed in Section 3 of the ESI†. Specifically, even under the most stringent conditions used in this report ($G \cong 0.5$ mL/s and $T_{cathode}$ = 170 °C), the temperature non-uniformity over most of the cathode surface was calculated to be within ca. 6 K (Fig. S3, ESI†), which is much smaller than the interelectrode temperature gap for these conditions (> 100 K, *vide infra*). Thus, we assume that the temperature measured by the thermocouple embedded in the cathode represents the temperature of the cathode surface with a spatial variance of up to ±3 K. The absence of degradation of the working liquid was confirmed by measuring an ultraviolet–visible (UV-vis) absorption spectrum after each experimental day and comparing it with that of the fresh liquid. Further details of the sample preparation, electrodes, experimental setup, measurements, and simulations are provided in Section 5.

## 3  Results and Discussion

### 3.1  Dependence of cooling properties on interelectrode channel geometry

First, the cooling properties are investigated and compared for the cases with cathode #1−3. Figure 3a compares the dependence of the heat removal ($Q$ [W]) on $T_{cathode}$ for $G$ = 0.52±0.03 mL/s.

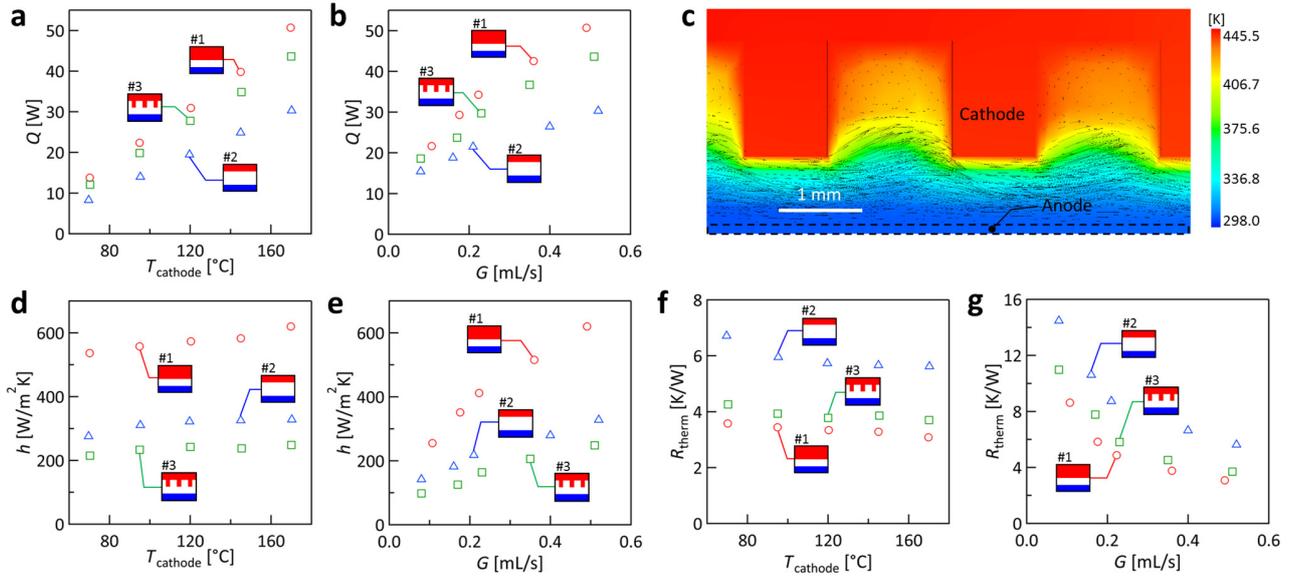

**Fig. 3** Cooling performance of the cell. Dependences of the heat removal on (a) $T_{cathode}$ at $G = 0.52\pm0.03$ mL/s and (b) $G$ at $T_{cathode} = 170$ °C. (c) Simulated temperature contour and liquid velocity vectors on the center cross-sectional plane of the cell with cathode #3 for $T_{cathode} = 170$ °C and $G = 0.49$ mL/s. Dependences of the heat transfer coefficient on (d) $T_{cathode}$ at $G = 0.52\pm0.03$ mL and (e) $G$ at $T_{cathode} = 170$ °C. Dependences of the cell thermal resistance on (f) $T_{cathode}$ at $G = 0.52\pm0.03$ mL and (g) $G$ at $T_{cathode} = 170$ °C.

The plots show that the $Q$ attained by cathode #1 was larger than that by cathode #2, which can be rationalized by the smaller hydraulic diameter ($D_h$; see Section 2 of ESI†) of the interelectrode channel formed by cathode #1 ($D_h = 1.54$ mm) than that by cathode #2 ($D_h = 4.0$ mm). However, different from our expectation, the use of cathode #3 with a heat transfer area twice that of the other cathodes resulted in a lower $Q$ than that of #1; this was also found for other $G$ (Fig. 3b). To understand this behavior, Fig. 3c shows the results of our simulation for cathode #3, displaying the temperature color contour and velocity vectors on the center cross-sectional plane parallel to the flow direction. It can be seen that the liquid in the valley between adjacent fins is almost stagnant and keeps circulating in the same valley, remaining separated from the main stream of the liquid. Because the liquid is almost stagnant in the valleys, it is considered to provide thermal resistance between the valley bottom and main stream of the liquid. This aspect may be attributed to the highly laminar flow ($Re < 3$) in the interelectrode channel in the present study. Among the cathodes tested, cathode #1 attained the highest $Q$ of 51 W at $T_{cathode}$ = 170 °C and $G$ = 0.49 mL/s.

To evaluate the cooling ability *per unit surface area*, the average heat transfer coefficient $h$ [W/(m²·K)] can be used. This is related to Newton's law of cooling[45,46]

$$Q = hA\left(T_{cathode} - T_{channel,in}\right), \qquad (3)$$

where $A$ is the heat transfer surface area and $T_{channel,in}$ is the liquid temperature at the entrance of the interelectrode channel obtained from our simulation (see Fig. S1b of ESI† for the position of $T_{channel,in}$). Although there is some freedom in the choice of the temperature difference used to calculate $h$, we chose $T_{cathode} - T_{channel,in}$ to exclude the minor temperature rise in the liquid between the cell entrance and entrance of the interelectrode channel. It should be noted that the choice of $T_{channel,in}$ for the lower temperature in eqn (3) results in a conservative calculation of $h$ compared to the case when the mean liquid temperature in the channel was used as the lower temperature in eqn (3). The values of $h$ were almost invariant with $T_{cathode}$ for all the cathodes (Fig. 3d) and increased with $G$ (Fig. 3e), which is typical for forced convection cooling.[45,46] The case with electrode #1 resulted in the highest $h$ and this is again attributed to its smaller $D_h$ compared with those of the cases with other electrodes. The case with electrode #3 yielded the smallest $h$, which can be explained by the stagnant flow in the valleys between fins found by the simulations (Fig. 3c) and the fact that $h$ is the evaluation measure *per unit surface area*.

Another useful measure is the thermal resistance[45,46] of the cell $R_{therm}$ [K/W] expressed by

$$R_{therm} = \frac{(T_{cathode} - T_{in})}{Q}, \qquad (4)$$

which evaluates the cooling performance of the whole unit (smaller is better). The term $T_{cathode} - T_{in}$ is used in eqn (4) because these temperatures are the fundamental conditions given. Lower $R_{therm}$ means that the temperature of a hot object is kept at lower temperature for the removal of the same $Q$. Figure 3f shows that $R_{therm}$ was almost constant over $T_{cathode}$ and decreased with increasing $G$. The order of $R_{therm}$ for the cases with different cathodes of #1 < #3 < #2 reflected the order of $Q$ in Figs. 3a and b.

Overall, the cell with cathode #1 showed the best cooling performance regardless of the evaluation measure used. This configuration attained an $h$ of 620 W/(m$^2$·K) at $G$ = 0.49 mL/s. As a comparison, $h$ in forced convection cooling using water ranges between 50 and 10,000 W/(m$^2$·K).[46] Considering the relatively low thermal conductivity of [C$_2$mim][NTf$_2$] (0.13 W/m·K at 300 K)[47] compared to that of water (0.61 W/m·K at 300 K)[46], this $h$ is believed to be reasonable.

### 3.2 Dependence of power generation properties on interelectrode channel geometry

Next, power generation properties for different interelectrode channels formed with cathode #1–3 are compared for fixed $T_{cathode}$ (170 °C) and $G$ (0.50±0.02 mL/s). Figure 4a compares the $I$- $V$ curves for the different cases. The case with cathode #2 resulted in the highest open-circuit voltage ($V_{oc}$) and lowest short-circuit current ($I_{sc}$). The highest $V_{oc}$ is ascribed to the largest interelectrode temperature difference $\Delta T$ ($\equiv T_{cathode} - T_{anode,ave}$) caused by the largest interelectrode distance in the case of #2. $T_{anode,ave}$ denotes the average surface temperature on the anode obtained by our simulations (see Section 5 for details). Specifically, $\Delta T$ ($T_{anode,ave}$) for the cases with cathode #1, 2, and 3 were found to be 131.5 K (35.7 °C), 137.1 K (31.3 °C), and 132.6 K (34.8 °C), respectively, supporting the above explanation of $V_{oc}$. The corresponding Seebeck coefficient calculated from $V_{oc}/\Delta T$ was 1.5–1.6 mV/K, which roughly agrees with the reported value of 1.64 mV/K in [$C_2$mim][$NTf_2$].[22,48] In contrast, the smallest $I_{sc}$ for the configuration with cathode #2 could be caused by the highest mass transfer resistance ($R_{mt}$) and/or solution resistance ($R_{sol}$) in the working liquid in the channel because this configuration had the largest interelectrode distance, which will be examined below.

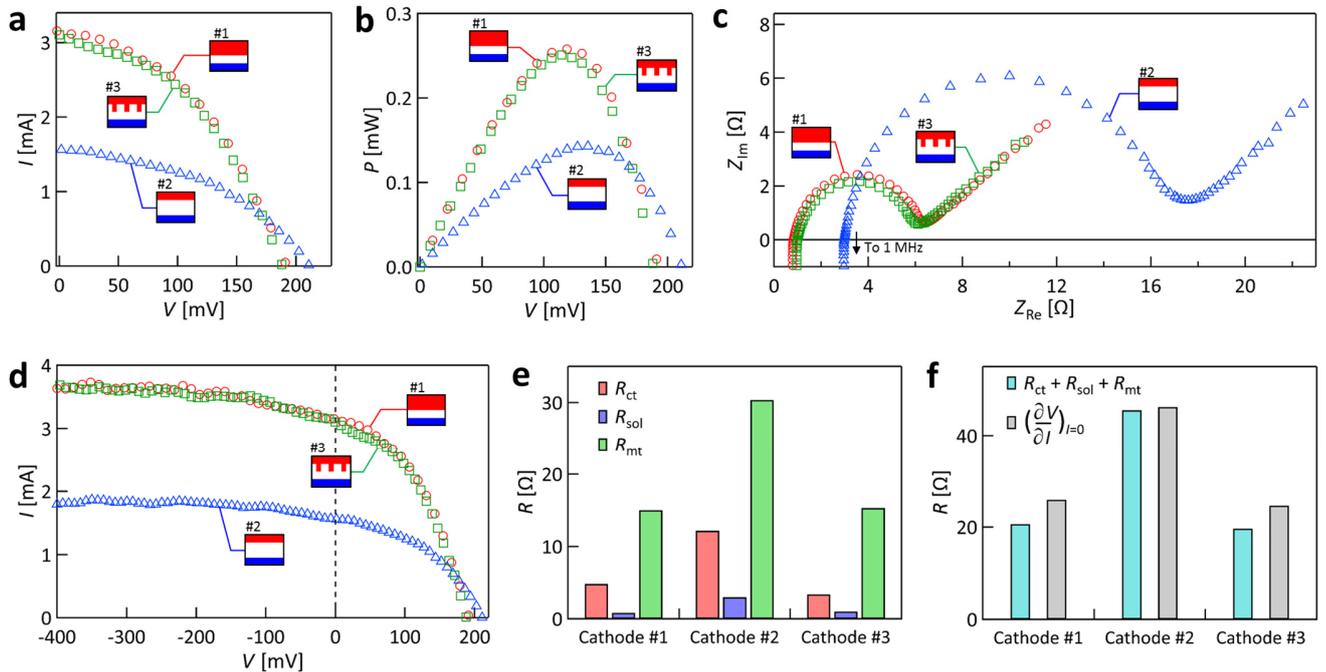

**Fig. 4** Dependence of the power generation characteristics on the cathode type for $T_{cathode}$ = 170 °C and $G$ = 0.50±0.02 mL/s. In these panels, the symbols representing interelectrode channels formed with cathode #1–3 introduced in Fig. 2c are used. (a) $I$-$V$ curves, (b) $P$-$V$ curves, and (c) Nyquist plots. (d) The same $I$-$V$ curves as those in panel (a) presented down to −400 mV. (e) Comparisons between the involved resistances. (f) Comparisons between the sum of $R_{ct} + R_{sol} + R_{mt}$ and the inverse slope of the $I$-$V$ curves at $I$ = 0.

Figure 4a also shows that the $I$-$V$ curves for the configurations with cathode #1 and 3 were similar to each other, which implies that not only $\Delta T$ but also the relevant electrochemical resistances were similar for these cases. Using the relation power ($P$) = $I \times V$, the $P$-$V$ curves are plotted in Fig. 4b. This shows that the cases with cathode #1 and #3 yielded maximum $P$ ($P_{max}$) of 0.26 and 0.25 mW, respectively; the case with cathode #2 was ca. 45% smaller because it had the smallest current.

To elucidate the electrochemical resistances of the cell equipped with different electrodes, AC impedance measurements were carried out during the same experiments. The Nyquist plots in Fig. 4c show semicircles that can be interpreted by the Randles equivalent circuit model.[49] Here, the diameter of the semicircle is assigned as the charge transfer resistance ($R_{ct}$) for the two electrodes in the cell. It is noted that $R_{ct}$ for one electrode could not be determined because of the asymmetric and non-isothermal two-electrode configuration used in this study; the diameter of the semicircle is regarded as the overall $R_{ct}$ for the two electrodes. The impedance at which the Nyquist plot intersects the real axis is assigned to $R_{sol}$ of the liquid in the interelectrode region.[49]

Mass transfer resistance is often important when characterizing cell properties. While its accurate evaluation is not straightforward, we may evaluate a small-signal mass transfer resistance ($R_{mt}$), which is for a mass transfer overpotential near the redox equilibrium related to the limiting current ($I_{lim}$).[49] According to the semi-empirical treatment under a steady-state diffusion layer approximation,[49] $R_{mt}$ is related to $I_{lim}$ by (see Section 4 of ESI† for the derivation)

$$R_{mt} \cong \frac{RT}{nF} \frac{2}{I_{lim}}, \tag{5}$$

where $R$ is the gas constant, $T$ is the absolute temperature, and $n$ is the number of electrons involved ($n = 1$ in the present case). Figure 4d shows the same $I$-$V$ curves as Fig. 4a extended into the negative $V$ region. All the curves saturated as $V$ was shifted to negative and we regard $I$ at −400 mV as $I_{lim}$. It should be noted that Abraham et al.[25] also calculated $R_{mt}$ for their stationary thermocell based on the same theoretical model but following a different route from ours (Section 4, ESI†).

Figure 4e compares $R_{ct}$, $R_{sol}$, and $R_{mt}$ for the cases with cathode #1–3, from which the following three points were noted. First, $R_{sol}$ was small for all the cases, which may be because of the use of ionic liquids that can work as electrolytes with high ion density.[38–40] The largest $R_{sol}$ for the cell with cathode #2 is ascribed to it possessing the largest interelectrode distance. Second, all the resistances of the cell equipped with cathode #1 were similar to those of the cell with cathode #3, which corroborates the similarity between their $I$-$V$ curves found in Fig. 4a. Third, $R_{mt}$ was dominant in all cases. In the cases with cathode #1 and #3, $R_{mt}$ contributed ca. 75% of the sum of $R_{ct} + R_{sol} + R_{mt}$. The dominance of $R_{mt}$ over other types of resistance was also reported previously for a stationary thermocell.[25]

Before proceeding further, we check the validity of our use of $R_{mt}$ by eqn (5) derived from the diffusion-layer approximation model[49] (Section 4 of ESI†). This validation was conducted because the experimental method used to obtain $R_{ct}$ and $R_{sol}$ (AC impedance measurements) and that used to obtain $I_{lim}$ in eqn (5) (steady state $I$-$V$ measurements) were different. In Fig. 4f, we compare the values of $(\partial V/\partial I)_{I=0}$ obtained from the slope of the $I$-$V$ curves on the horizontal axis of Fig. 4a and the sum of $R_{ct} + R_{sol} + R_{mt}$. Here, the former values are considered to represent the total resistance near equilibrium. The agreement between the $(\partial V/\partial I)_{I=0}$ values and the sum of $R_{ct} + R_{sol} + R_{mt}$ was satisfactory for all the cases (within 20%). Thus, we assumed that eqn (5) can be used to estimate $R_{mt}$ in the present cell.

From the results of Figs. 3 and 4, we consider that the use of cathode #1 is the best in the sense that it simultaneously achieved higher cooling and power generation performances. Below, we further investigate the cell properties using cathode #1.

### 3.3 Dependence of power generation properties on cathode temperature

Figure 5 shows the effect of $T_{cathode}$ on the power generation properties examined using cathode #1 for $G = 0.50\pm0.01$ mL/s. The $V \geq 0$ region of the $I$-$V$ curves (Fig. 5a) reveals that both $V_{OC}$ and $I_{SC}$ increased with rising $T_{cathode}$. The former is attributed to the increase of $\Delta T$ with $T_{cathode}$. The latter could be caused by the decrease of the mean viscosity of the liquid in the interelectrode channel, which has been estimated by our simulation to be −30% for the change of $T_{cathode}$ from 70 to 170 °C. However, this estimated decrease is minor compared to the observed increase of $I_{SC}$ by a factor of three caused by this change in $T_{cathode}$ (Fig. 5a), which will be discussed below.

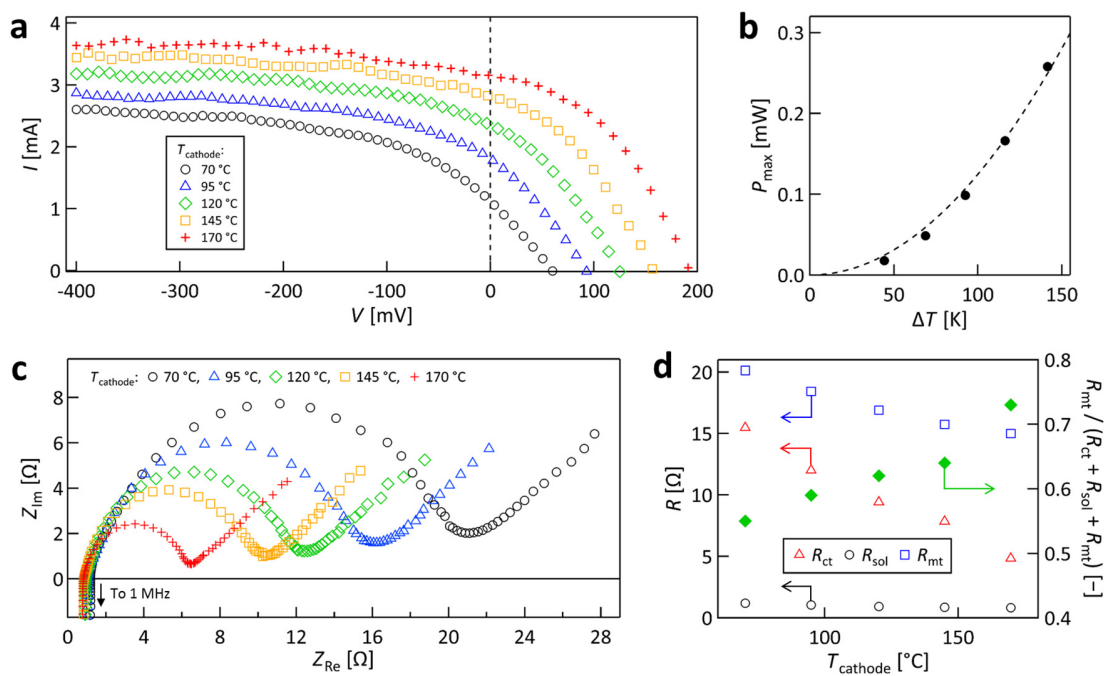

**Fig. 5** Dependence of the power generation characteristics on the cathode temperature for cathode #1 and $G = 0.50\pm0.01$ mL/s. (a) $I$-$V$ curves down to −400 mV. (b) Plot of $P_{max}$ vs. $\Delta T$ with a quadratic fit (dashed curve). (c) Nyquist plots. (d) Dependence of the resistances (left axis) and the fraction of $R_{mt}$ over the sum of $R_{ct} + R_{sol} + R_{mt}$ (right axis) on the cathode temperature.

One characteristic noticed in Fig. 5a was that the shape of the $I$-$V$ curve for $V \geq 0$ changed with $T_{cathode}$. This can be explained using the results of Fig. 5a, which revealed that a higher (lower) $T_{cathode}$ resulted in faster (slower) saturation of $I$ as $V$ was lowered from $V_{OC}$ to −400 mV. That is, the $I$-$V$ curve at $T_{cathode}$ = 70 °C was straight or Ohmic-like[15] for $V \geq 0$ because $I$ near $V = 0$ was still far from $I_{lim}$ imposed by the mass transfer; i.e., $I$ for $V \geq 0$ was considered to be more limited by other factors than by mass transfer. In contrast, the $I$-$V$ curve at $T_{cathode}$ = 170 °C was bent at $V \geq 0$ because $I$ near $V = 0$ was already close to $I_{lim}$; i.e., this bent shape at $V \geq 0$ mostly arose from the mass transfer limitation. These $I$-$V$ curves could be fitted by the model used for eqn (5) (Fig. S6, ESI†). Abraham et al.[25] fitted their $I$-$V$ curves differently using the same model.

Figure 5b shows the dependence of $P_{max}$ on $\Delta T$. The quadratic increase may be the manifestation of relations of $V \propto \Delta T$ and power = $V^2$/resistance; however, the resistance itself should depend on the temperature. To elucidate the temperature dependence of the different types of resistance, AC impedance measurements were carried out during these experiments. $R_{ct}$ and $R_{sol}$ were obtained from the Nyquist plots (Fig. 5c) and $R_{mt}$ was calculated using $I_{lim}$ (Fig. 5a) as described above.

Figure 5d plots the dependences of $R_{ct}$, $R_{sol}$, and $R_{mt}$ on $T_{cathode}$. Again, $R_{sol}$ was minor and $R_{mt}$ was dominant for all the $T_{cathode}$ tested. The decreases of $R_{mt}$ and $R_{sol}$ with increasing $T_{cathode}$ are ascribed to the decrease of the mean liquid viscosity in the channel with rising temperature. Quantitatively, $R_{mt}$ decreased by ca. 25% as $T_{cathode}$ rose from 70 to 170 °C, which is close to the aforementioned change of the mean liquid viscosity of −30% for this change of $T_{cathode}$. On the right axis, the fraction of $R_{mt}$ over $R_{ct} + R_{sol} + R_{mt}$ is plotted, indicating that the relative importance of $R_{mt}$ increased with $T_{cathode}$. This is consistent with the above explanation for the change in the shape of the I-V curves at $V \geq 0$ observed in Fig. 5a.

As shown in Fig. 5d, while $R_{ct}$ was ca. 20% smaller than $R_{mt}$ at $T_{cathode}$ = 70 °C, it was ca. 70% smaller than $R_{mt}$ at 170 °C. To evaluate this rapid decrease of $R_{ct}$ with increasing $T_{cathode}$, we considered the work of Tachikawa et al.,[16] who investigated the temperature-dependent electrode kinetics of iron complexes in ionic liquids based on the theoretical framework of solvent reorganization dynamics.[50] According to their analyses, the apparent activation energy for the rate constant of the outer-sphere redox reaction on an electrode surface, $k_0$ ($\propto R_{ct}^{-1}$), is the sum of the relevant reorganization energy ($\Delta G^{\ddagger}$) and activation energy of the solvent viscosity ($E_{a(\eta)}$). They proposed that $\Delta G^{\ddagger}$ was minor compared to $E_{a(\eta)}$ and thus the apparent activation energy was similar to $E_{a(\eta)}$.[16] In the present study, from the temperature dependence of the working liquid viscosity (Fig. S7, ESI†), $E_{a(\eta)}$ was found to be 23.5 kJ/mol (Fig. S8a, ESI†) and the activation energy for the electrode kinetics of $R_{ct}^{-1}$ was found to be 20.5 kJ/mol (Fig. S8b, ESI†). Thus, both activation energies are similar to each other. Therefore, we consider that the present electrode kinetics (the dependence of $R_{ct}$ on $T_{cathode}$ in Fig. 5d) followed a similar mechanism to that proposed by Tachikawa and colleagues.[16] Back to the discussion of Fig. 5a above, the observed increase of $I_{SC}$ by a factor of three with the change of $T_{cathode}$ from 70 to 170 °C may partially be attributed to this rapid decrease of $R_{ct}$ with rising temperature.

### 3.4 Dependence of power generation properties on liquid flow rate

Figure 6 shows the effect of $G$ on the power generation properties investigated for the cell with cathode #1 at $T_{cathode}$ = 170 °C. The $I$-$V$ curve (Fig. 6a) reveals that $V_{OC}$ increased and $I_{lim}$ ($I$ at –400 V) decreased with increasing $G$. The increase of $V_{OC}$ is attributed to the decrease of $T_{anode,ave}$ with increasing $G$, as supported by our simulation (Fig. S9, ESI†). The decrease of $I_{lim}$ with rising $G$ is ascribed to the increase in the liquid viscosity induced by the lowering of the mean temperature in the liquid channel, which will be examined below.

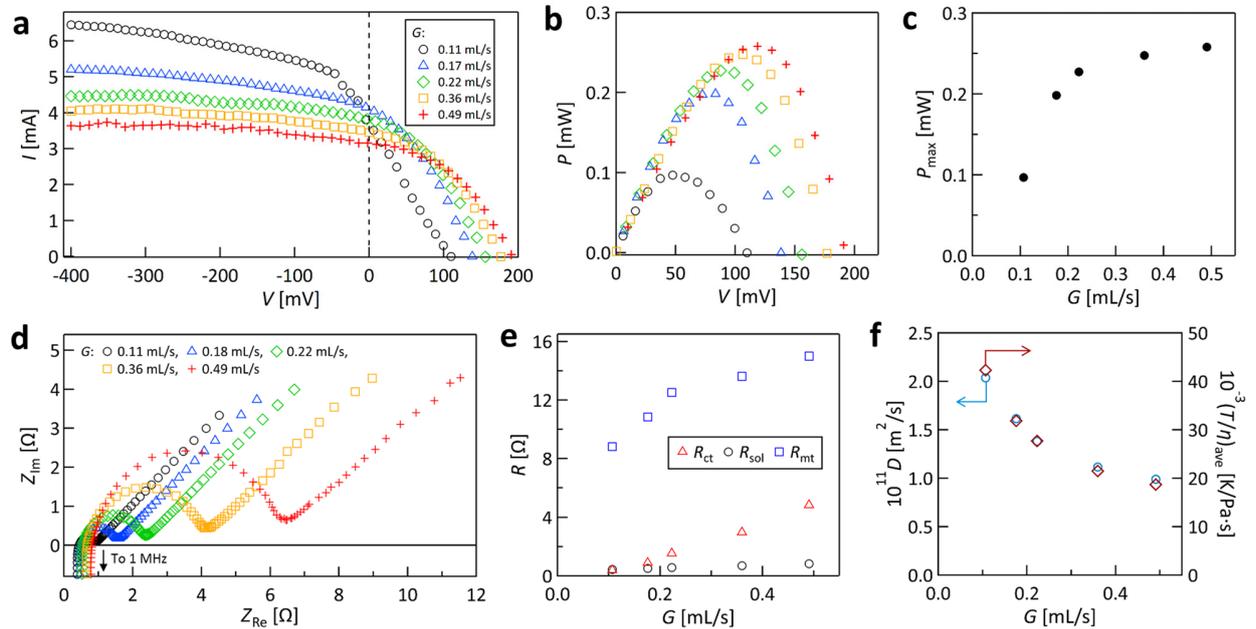

**Fig. 6** Dependence of the power generation characteristics on flow rate ($G$) for the cell with cathode #1 and $T_{cathode}$ = 170 °C. (a) $I$-$V$ curves. (b) $P$-$V$ curves. (c) Plot of $P_{max}$ vs. $G$. (d) Nyquist plots. (e) Dependence of resistances on $G$. (f) Dependences of $D$ calculated using eqn (7) (left axis) and volumetric average of $T/\eta$ in eqn (8) (right axis) on $G$.

Figure 6b shows the *P-V* curves, from which $P_{max}$ for each *G* was obtained. Figure 6c plots $P_{max}$ vs. *G*. $P_{max}$ increased remarkably with rising *G* when *G* was low (< 0.25 mL/s), which is mainly attributed to the increase of the interelectrode $\Delta T$ or $V_{OC}$ with *G*, as seen in Fig. 6a. However, for *G* > 0.25 mL/s, the increase of $P_{max}$ slowed down and became nearly saturated. This was ascribed to the nearly balanced decrease in *I* and increase in *V* with rising *G*, as indicated in Fig. 6a. Therefore, in the present cell, the enhancement of $P_{max}$ with increasing *G* was weakened by the concomitant decrease in *I*.

These findings were further examined by electrochemical analyses. Figure 6d shows the Nyquist plots acquired during the same experiments. $R_{ct}$, $R_{sol}$, and $R_{mt}$ were obtained as before. Figure 6e plots the dependence of these resistances on *G*. Similar to the findings described above, $R_{mt}$ was dominant and $R_{sol}$ was small, while $R_{ct}$ was between them. The increases of $R_{mt}$ and $R_{ct}$ with increasing *G* are consistent with the decrease of $I_{lim}$ found in Fig. 6a.

We now test our above explanation in terms of the diffusion of the redox species in the channel. The diffusion constants of the oxidized species ($D_O$) and reduced species ($D_R$) are related to the slope ($\sigma$) of a plot of impedance vs. (frequency)$^{-1/2}$, called a Randles plot, by the following relation[49]

$$\sigma = \frac{RT}{F^2 A \sqrt{2}} \left( \frac{1}{D_O^{1/2} C_O^*} + \frac{1}{D_R^{1/2} C_R^*} \right), \tag{6}$$

where *A* is the electrode surface area and $C_O^*$ and $C_R^*$ are the concentrations of oxidized and reduced species in the bulk electrolyte, respectively. The Randles plots generated from the data in Fig. 6d are presented in Fig. S10 of ESI†. Generally, $D_O$ and $D_R$ in ionic liquids are not the same.[33] However, here we assumed $D_O \cong D_R \equiv D$ as a first approximation (see also Fig. S5 of ESI† for the forward and backward limiting currents measured under the same $T_{cathode}$ and *G*). Then, eqn (6) can be rewritten as

$$D = 2\left(\frac{RT}{C\sigma F^2 A}\right)^2. \tag{7}$$

The left axis of Fig. 6f plots $D$ calculated using eqn (7) against $G$. The resulting relationship supports the viewpoint that the diffusional mobility of the redox species decreased as $G$ increased.

To further test this idea from a theoretical viewpoint, we used the Stokes–Einstein equation,[51] which relates $D$ to the solvent viscosity ($\eta$) as follows

$$D = \frac{k_B T}{6\pi \eta r} \propto \frac{T}{\eta}, \tag{8}$$

where $k_B$ is the Boltzmann constant and $r$ is the solute radius. We define $(T/\eta)_{ave}$ as the volumetric average of $T/\eta$ in the interelectrode channel; values for $(T/\eta)_{ave}$ were obtained from our simulations. The right axis of Fig. 6f plots $(T/\eta)_{ave}$ against $G$. The good agreement between the decrease of $D$ experimentally obtained using eqn (7) on the left axis and $(T/\eta)_{ave}$ that appeared theoretically in eqn (8) supports the above interpretation of the decrease of $I$ (Fig. 6a) and increase of $R_{mt}$ (Fig. 6e) with increasing $G$.

### 3.5 Measures to evaluate cell performance

So far, we have regarded the cell with cathode #1 as the best case in the sense that it simultaneously attained the largest $Q$ and $P_{max}$ among all the cathodes tested. Besides this perspective, consideration of the following three points is thought to be beneficial for this type of thermocell where the role of electric power generation has been integrated into forced convection cooling.

The first point is that pumping work is necessary for this forced-flow cell. The hydrodynamic pumping work $W_{pump}$ required to force the liquid through the cell it is given by $G$ (in units of m³/s) times the pressure drop between the cell entrance and exit $\Delta P$ as

$$W_{pump} = G\Delta P. \tag{9}$$

This $W_{pump}$ does not include mechanical friction in the pump and pumping work for other parts of the fluid loop, and this ideality is like the Carnot efficiency, which is often used as a reference to the ideal limit.[12,21,32] In this regard, we should also consider that the cell with cathode #1 has the narrowest interelectrode channel (cf. Fig. 2c) and thus would require the largest $\Delta P$ and $W_{pump}$ of the configurations with different cathodes for the same $G$; the inclusion of this point into our assessment will be attempted below.

Second, the efficiency ($\phi$) conventionally used to assess solid thermoelectric cells and stationary liquid thermocells is defined by the ratio of the generated power to the heat input ($\phi \equiv P_{max}/Q$). If we apply this $\phi$ to the present system, $\phi$ for the configurations with cathode #1, 2, and 3 at $T_{cathode}$ = 170 °C and $G \cong 0.5$ mL/s are $5.1 \times 10^{-6}$, $4.7 \times 10^{-6}$, and $5.8 \times 10^{-6}$, respectively. However, $\phi$ is not considered to be a suitable parameter for evaluating the present system. This is because the use of $\phi$ as a performance measure would lead to the conclusion that the cell with cathode #1, which resulted in a similar $P_{max}$ and higher $Q$ than those of the cell with cathode #3 (Figs. 4 and 3, respectively), was inferior to that with cathode #3 by a factor of 5.1/5.8 due to the better cooling performance or $Q$, which is in the denominator of $\phi$, of the former compared with that of the latter. The use of $\phi$ thus contradicts the primary purpose of the present system.

Third, in solid-state thermoelectric cells[52,53] and stationary liquid thermocells,[23,26] almost all heat input from the hot-side electrode passes through the cell and reaches the counter cold-side electrode. In such cases, $Q$ may be regarded as an unconditionally available energy resource because active cooling of the hot-side plane is not intended, rendering $\phi$ an appropriate parameter for the performance evaluation. Conversely, in the present forced-flow cell, only a very small fraction of $Q$ departed from the hot electrode reaches the cold electrode because most $Q$ is taken away by the liquid flowing through the channel. Our simulations using $T_{cathode}$ = 170 °C and $G \cong$ 0.5 mL/s revealed that the heat reaching the anode was only 0.4±0.05 W for all the cases with different cathodes, which is ca. 1% of the $Q$ departed from the cathode. Therefore, the heat flow behavior in the present forced-flow cell is different from that in the aforementioned conventional cells; this reflects the fact that $Q$ in the present study is regarded as undesirable and has to be promptly removed (cf. Section 1) rather than an unconditionally utilizable energy source.

Considering these points, below we introduce some measures that may be suitable for evaluating such forced-flow thermocells. The first is the *dimensionless gain* ($\Lambda$), which is defined as the ratio of the generated power to $W_{\text{pump}}$

$$\Lambda = \frac{P_{\max}}{W_{\text{pump}}}. \tag{10}$$

If $\Lambda$ is larger than unity, excess electric work is generated by the cell besides the hydrodynamic pumping work consumed to force the coolant liquid through the cell. In this report, $\Delta P$ was obtained from the simulations because our cell was so small that it was difficult to conduct reliable experimental measurements; in contrast, our simulations were quantitatively reliable (Section 5) because of the highly laminar conditions used.

Figure 7a plots $\Lambda$ vs. $G$ for $T_{\text{cathode}} = 170$ °C, where the actual values of $G$ in corresponding experiments have been used on the horizontal axis. For all the cases, $\Lambda$ and $G$ are inversely correlated and $\Lambda$ exceeded unity for $G$ below ca. 0.36 mL/s. Upon further lowering $G$, $\Lambda$ increased with the sacrifice of the cooling performance (cf. Fig. 3b). The rapid increase of $\Lambda$ as $G \to 0$ was because $W_{\text{pump}} \to 0$ in eqn (10) at this limit. Figure 7a reveals that the relations between $\Lambda$ and $G$ for all the cases with different cathodes fall on almost the same curve. This tendency could be interpreted that the higher $P_{\max}$ obtained using cathode #1 was realized at the cost of the larger pumping work to force the liquid through its narrower interelectrode channel. $\Lambda$ became lower than unity for $G > 0.36$ mL, which can be ascribed to the saturation of $P_{\text{max}}$ at high $G$ (cf. Fig. 6c). Our simulations revealed that more than half of $\Delta P$ was caused by the narrow feed holes with a diameter of 2 mm near the entrance and exit of the cell (cf. Fig. S1a, ESI†). Therefore, decreasing the flow resistances in these feed holes would further enhance $\Lambda$.

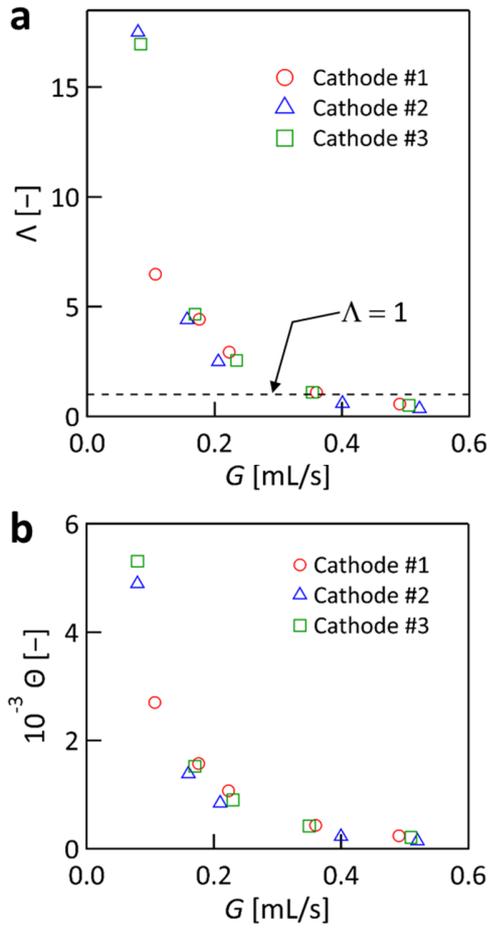

**Fig. 7** Plots of (a) $\Lambda$ and (b) $\Theta$ calculated using eqn (10) and (11), respectively, against flow rate ($G$) for the cell equipped with cathode #1−3 at $T_{\text{cathode}} = 170$ °C.

Although $\Lambda$ includes the pumping work needed for the cell, it does not include a factor to evaluate cooling ability. To amend this point, another dimensionless number is introduced,

$$\Theta = \left(\frac{P_{\max}Q}{W_{\text{pump}}^2}\right)^{\frac{1}{2}}. \tag{11}$$

Larger $\Theta$ is better both in terms of cooling and power generation. Figure 7b plots $\Theta$ vs. $G$ for $T_{\text{cathode}}$ = 170 °C. $\Theta$ exhibited a similar tendency to that of $\Lambda$ in Fig. 7a. Again, all the plots fall on almost the same curve. This result may also be interpreted that better cooling *and* power generation performances are attained at the cost of larger $W_{\text{pump}}$. However, each specific value of $Q$, $P_{\max}$, and $W_{\text{pump}}$, is strongly influenced by the geometry of the interelectrode channel, as shown above. The confirmation of $\Lambda > 1$, although the range of $G$ was limited, would justify the concept of such forced-flow thermocells. Further improvements may be achieved by refined design of the cell geometry, choice of electrode materials, and the use of novel high-density redox-couple liquids that may emerge in the future.

## 4 Conclusions

Motivated by the unaddressed issue of the loss of the exergy that presently accompanies many situations of forced convection cooling, we designed a forced-flow thermocell in which an electrolyte liquid flowed through an interelectrode channel where the cathode simulated a heat-emitting plane that needed to be cooled. Throughout the experiments, the cell with cathode #1, which formed the narrowest channel, exhibited the best cooling performance under all the $G$ and $T_{\text{cathode}}$ conditions tested, which agreed with the prediction by the classical heat-transfer theory. Different from our expectation, cathode #3 with a fin-like extended heat-transfer surface displayed poorer cooling ability than cathode #1. This was caused by the stagnant flow in the valleys between fins, which probably originated from the highly laminar conditions in the present study. The power generation characteristics of a cell with different cathodes were also compared and in all cases $R_{\text{mt}}$ was the dominant factor limiting the current. $R_{ct}$ decreased rapidly with increasing $T_{cathode}$, which may result from the electrode kinetics being mostly controlled by the activation energy of the electrolyte viscosity. The increase of $P_{max}$ was found to saturate with increasing $G$. This was caused by the increase of the resistances with $G$, which was attributed to the decrease of the diffusion coefficient of the redox couple in the liquid with increasing $G$.

To appropriately consider the purpose of such forced-flow thermocells, we introduced a dimensionless gain ($\Lambda$) and its modified number ($\Theta$). For $G < 0.36$ mL/s, $\Lambda$ was above unity, demonstrating that the cell can generate a larger electric power than the hydrodynamic pumping work required to force the coolant through the cell. This result supported the concept of such kind of thermocells. We noticed, however, that the discussion regarding the introduction of $\Lambda$ and $\Theta$ may still be premature and requires further validation. Therefore, in the meantime, simultaneous use of multiple evaluation measures, i.e., $Q$, $R_{therm}$ (and/or $h$), $\phi$, and $\Lambda$ (and/or $\Theta$), would be recommended.

## 5 Methods

**5.1 Chemicals**

The solvent ionic liquid [C$_2$mim][NTf$_2$] (Fig. 1a) was purchased from IoLiTec, Germany. The certified purity and water content were > 99.5% and < 100 ppm, respectively. The high purity of the received [C$_2$mim][NTf$_2$] was confirmed by the absence of visible-range absorption and sufficiently weak UV absorption (Fig. S11, ESI†). The temperature dependence of the viscosity of this ionic liquid was measured and the result agreed well with previously reported values (Fig. S12, ESI†).

The redox reagents Co$^{II}$(bpy)$_3$(NTf$_2$)$_2$ and Co$^{III}$(bpy)$_3$(NTf$_2$)$_3$ (Fig. 1b) were supplied by Nippon Kayaku, Japan and their certified purity was > 98%. This good redox couple was developed by the Pringle-MacFarlane team and has been widely used in thermoelectrochemical conversion because of its high thermal and air stabilities and large Seebeck coefficient.[22] To check the reliability of the chemicals we received, we compared the optical absorption spectrum of the solution with that of the same redox couple purchased from a different supplier (Dyenamo, Sweden). The agreement between these optical absorption spectra was excellent regarding both spectral shape and quantitative absorbance (Fig. S13, ESI†), indicating that the redox reagents Co$^{II/III}$(bpy)$_3$(NTf$_2$)$_{2/3}$ used in this report were of high quality.

**5.2 Preparation and properties of the working liquid**

The preparation, handling, and storage of the working liquid (0.06 M solution of Co$^{II/III}$(bpy)$_3$(NTf$_2$)$_{2/3}$ in [C$_2$mim][NTf$_2$]) were carried out in a glovebox equipped with a circulation gas purifier (OMNI-LAB, Vacuum Atmospheres Company). In this glovebox, the moisture and oxygen concentrations were kept below 1 ppm. First, as-received [C$_2$mim][NTf$_2$] was dried under vacuum at 110 °C for 3 h in an open glass vessel with the liquid surface height below ca. 5 mm to facilitate the escape of moisture. The powders of Co$^{II}$(bpy)$_3$(NTf$_2$)$_2$ and Co$^{III}$(bpy)$_3$(NTf$_2$)$_3$ were then dissolved in [C$_2$mim][NTf$_2$] in an open-top glass bottle by stirring the mixture on a hot-plate stirrer at 50 °C for 15 h. This process was conducted in the glovebox to further remove moisture from the liquid. The complete dissolution of the redox couple was checked by irradiating the liquid with a laser beam (wavelength: 632.8 nm), which confirmed the absence of light scattering from powder particles. This check was also performed before and after each experiment to confirm that the redox species did not precipitate from the solution. The working liquid was stored in the dark in the glovebox. Just before the experiment, the liquid was degassed in a vacuum oven at 50 °C for about 40 min. We found that this degassing treatment was important to prevent the emergence of gas bubbles during experiments.

### 5.3  Liquid circulation and flow rate measurement

The working liquid was circulated using a roller tubing pump (Masterflex, Cole-Parmer) equipped with a PTFE-tubing pump head (HV-77390-00, Cole-Parmer). The liquid flow rate was modulated by the rotation speed of the head. The actual flow rate or $G$ was measured for each experiment by introducing an air plug into the loop from a manifold junction in the middle of the loop and recording a digital movie to determine the progression rate of the air plug.

### 5.4  Fabrication and treatment of the electrodes

The anode was processed from a Pt plate (thickness: 0.3 mm; purity: > 99.98%) to dimensions of 29.5 × 25 mm for the directions parallel and perpendicular to the flow, respectively, by discharge cutting using an electrical wire. Cathode #1–3 (Fig. 2c) were fabricated by machine processing of a nickel block (purity: > 99.8%). The surfaces of the cathodes were sputter-coated with a Pt layer with a thickness of 100–130 nm to form a reversible electrode. Before each experiment, the anode was polished using a water slurry of alumina powder, rinsed with ultra-high purity water with a resistance of 18.2 MΩ·cm, and dried well in an oven at 70 °C. After each series of experiments, the cell was deconstructed and the electrodes were carefully cleaned by repeated ultrasonication first in acetone and then in methanol. Using these electrode treatments, we attained data reproducibility within 5%.

### 5.5  Heater and temperature control

The temperature of the cathode was set at a target value using a ceramic heater (BVP-004, Bach Resistor Ceramics GmbH, Germany) tightly contacted with the top plane of the cathode (cf. Fig. 2b). The heater body was composed of silicon nitride with dimensions of 18 × 18 × 3 mm. A sheathed thermocouple with a diameter of 1 mm was embedded inside the ceramic heater and this output was used for fast feedback control at a frequency of 50 Hz. The electric power consumption by the heater was measured in each experiment using a power meter (Model 3333, Hioki, Japan). This value was used to calculate the amount of $Q$ (see below). $T_{cathode}$ was measured using a teflon-coated T-type thermocouple. This junction was embedded inside a hole with a diameter of 1.4 mm and depth of 2.5 mm drilled in the top plane of the cathode. The hole was located on the centerline and near the downstream end of the cathode (see Fig. S3 in ESI† for the exact position). The thermocouple junction was fixed at the bottom of the hole filled with a heat-conductive resin.

### 5.6 Setup of the cell in the holder

As shown in Fig. 2d, the cell was suspended in the air using a holder made of aluminum to reproducibly suppress the physical contact between the cell unit and the surrounding environment. The cell unit was lifted up from the inner bottom of the "holder (bottom)" by a few millimeters using four PTFE screws ("lifting screws" in the side view of Fig. 2e) that penetrated from the bottom of the "holder (bottom)". The "holder (top)" was used to apply a downward force onto the thermal insulator and the ceramic heater so that the latter closely contacted the top plane of the cathode. By tightening the three plastic screws ("tightening screws" in Figs. 2d and e) that penetrated the "holder (top)" plate, a downward force was applied constantly to the thermal insulator, affording a thermal contact between the heater and cathode during experiments.

### 5.7 Electrical measurements

The *I-V* curves were obtained using an SMU (Model 2450, Keithley). All scans were started from open-circuit conditions. Details of the parameter settings of the SMU are given in Section 12 of ESI†. The AC impedance measurements between the cathode and anode were conducted using a potentiostat–galvanostat (VersaSTAT4, Princeton Applied Research). In each measurement, a sinusoidal input with an amplitude of 5 mV around $V_{oc}$ was applied and the AC frequency was scanned from 1 MHz to 0.1 Hz. We found that a waiting time of 30 min after the last change of the experimental conditions ($T_{cathode}$ or $G$) was sufficient to attain a steady state. However, to further ensure that a steady state was reached, at least 45 min was allowed to pass after the last change of the experimental conditions before conducting measurements.

### 5.8 Numerical simulations

Simulations were carried out using ANSYS Fluent® version 18.1. Tetrahedral meshes were used for the solid and fluid body and surface inflation meshes with four layers were set on solid surfaces (cf. Fig. S1, ESI†). The inflation meshes were used to treat the boundary layers on the solid surfaces. To avoid numerical difficulties, the highest value of the mesh skewness in the tetrahedral meshes was suppressed below 0.90 in all the simulations. The gravity (buoyancy) effect on the liquid and the heat escape from the surfaces to the ambient environment by both natural convection and thermal radiation were included in all the simulations. The temperature dependence of the viscosity of the working liquid was also included using the Vogel–Fulcher–Tammann parameters experimentally determined from our measurements of the temperature dependent viscosity (Fig. S7, ESI†). All the calculated surface temperatures of the cell unit agreed satisfactorily (within 5 K) with those experimentally measured using the thermocouples (see Fig. 2b for the positions).

### 5.9 Calculation of heat removed by the working liquid

The rate of heat removal by the working liquid $Q$ was calculated from the power consumption of the ceramic heater measured by the power meter (Model 3333, Hioki) minus the amount of the heat escaped from the cell surface to the ambient air. The latter was obtained using a pre-calibrated relationship between the surface temperatures and the steady-state power consumption in the ceramic heater with $G = 0$ (i.e., no liquid flow). It should be noted that we did not calculate $Q$ using $Q = \rho \times G \times C \times (T_{out} - T_{in})$, where $\rho$ and $C$ are the density and specific heat of the working liquid, respectively. Although this process also yielded quantitatively similar values to those obtained by the aforementioned method, we did not adopt the latter approach because we found that the multiplication of the four factors on the right-hand side of this relation caused intolerably large errors.

## Conflicts of interest

There are no conflicts of interest to declare.

## Acknowledgements

This work was financially supported by TEPCO Memorial Foundation Research Grant (Basic Research, Grant# H26(26)). We cordially thank Prof. Hiroshi Segawa (Univ. Tokyo), Dr. Takurou Murakami (AIST), Prof. Koichi Hishida (Keio Univ.), and Prof. Ichiro Yamanaka (Tokyo Tech) for valuable discussions, and Mr. Kazuki Niimi and Ms. Noriko Kiyoyanagi (Nippon Kayaku Co., Ltd.) for high purity redox reagents used in this study. We also thank Mr. Katsuhiko Otsuyama (Tokyo Tech) for his assistances in the design and fabrication of the experimental system used in this study and Dr. Natasha Lundin for correcting the English used in this report.

**Electronic Supplementary Information**

# Integration of Liquid Thermoelectrochemical Conversion into Forced Convection Cooling

Yutaka Ikeda, Kazuki Fukui, and Yoichi Murakami*

School of Engineering, Tokyo Institute of Technology, 2-12-1-I1-15 Ookayama, Meguro-ku, Tokyo 152-8552, Japan

*Corresponding Author: Yoichi Murakami, murakami.y.af@m.titech.ac.jp

**List of Contents**

**1. Cross-sectional view of the computational graphic of the cell**

Figure S1 shows the cross-sectional graphic of the cell used in this report with the computational meshes employed in our simulations for cathode #1. This computational graphic reflects the actual structure and dimensions of the present cell. Figure S1a also shows the position of the thermocouple hole in the cathode. Figure S1b indicates the position of $T_{channel,in}$. Figure S1c shows the color-contour map on the center cross section obtained from our simulation with $T_{cathode}$ = 170 °C and $G$ = 0.5 mL/s. Details of our simulations are given in Section 5 of the main text.

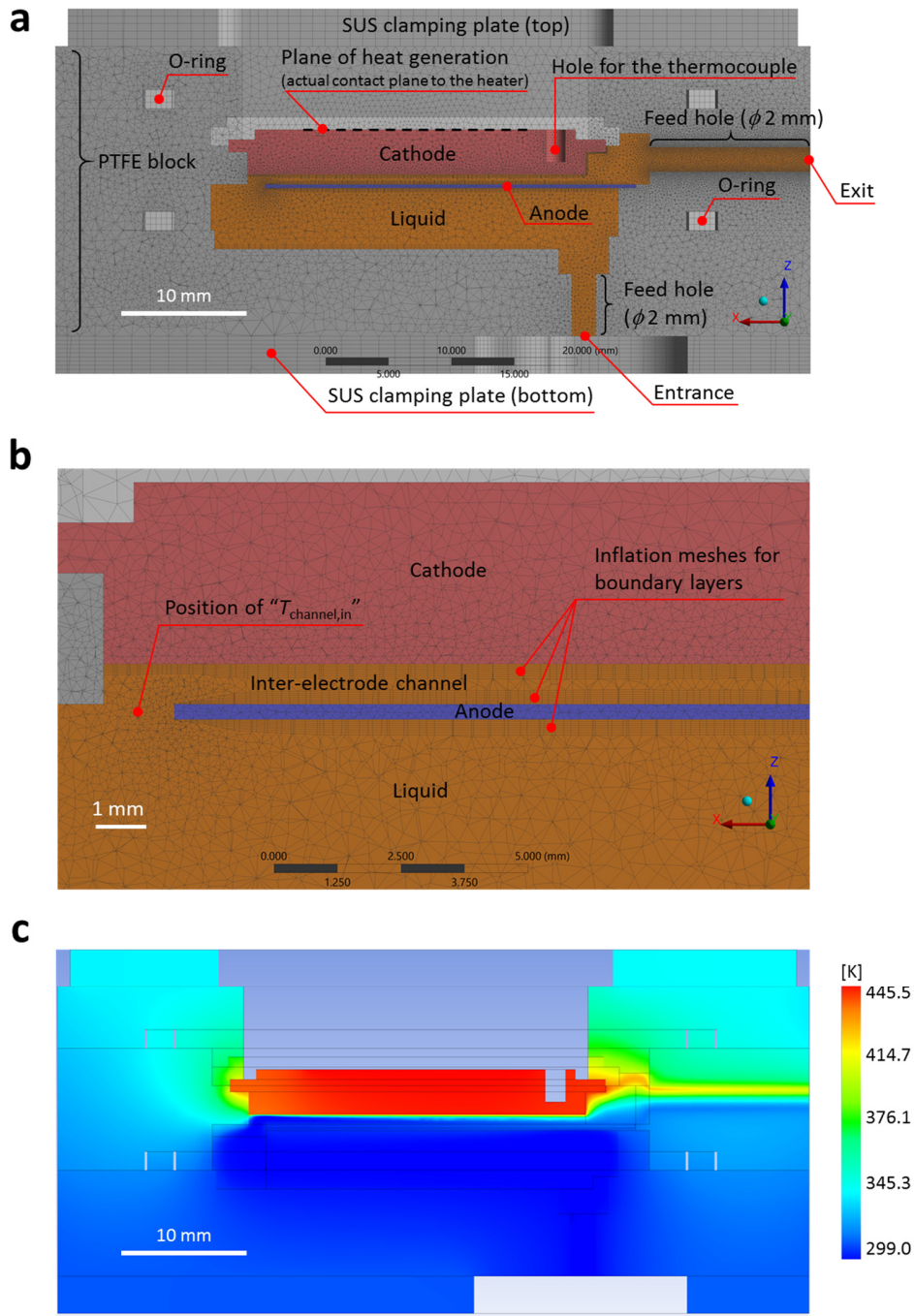

**Fig. S1** Graphic of the cross section of the cell unit showing the meshes used in our simulations for cathode #1. This cross section is on the center plane of the cell cut parallel to the direction of the liquid flow. (a) The whole cross-section, (b) a magnified view around the entrance of the interelectrode channel, and (c) the temperature color-contour plot on the same cross section for $T_{\text{cathode}}$ = 170 °C and $G$ = 0.5 mL/s. In panel (b), the position of $T_{\text{channel,in}}$ is indicated.

## 2. Theoretical basis for the use of narrow channel widths

A hydraulic diameter ($D_h$) is often used in evaluations of convection heat transfer.[S1,S2] $D_h$ of a rectangular duct with cross-sectional dimensions of $a \times b$ is $2ab/(a + b)$.[S1] In the case of forced convection cooling using a laminar flow in a channel, the heat transfer coefficient $h$ [W/m²·K] is inversely proportional to $D_h$ of the channel. This is because the Nusselt number ($Nu = hD_h/k$, where $k$ is the thermal conductivity of fluid)[S1] for a fully developed flow is almost constant (usually somewhere between 3.5 and 4.5) regardless of the boundary conditions and the channel cross-sectional geometry.[S1,S2] Therefore, the easiest method to enhance $h$ is to decrease $D_h$; this is the concept used in microchannel heat sinks, as mentioned in the main text.

A high $h$ corresponds to a thin thermal boundary layer on a solid wall, which is a spatially steep temperature gradient in the liquid flowing along the wall.[S1] Such thin thermal boundary layers are beneficial for establishing a large steady-state temperature difference between two closely spaced electrodes in the present study (cf. Fig. 2 in the main text). For the same liquid flow rate ($G$), the use of cathode #1 results in a higher liquid velocity in the interelectrode channel than that in the case when cathode #2 is used because of the smaller cross-sectional area of the former than that of the latter, which also contributes to the formation of a thin thermal boundary layer in the case of cathode #1. Therefore, the use of cathode #1 is expected to result in a higher $h$ and thus a higher heat removal rate than the use of cathode #2; this expectation was proven by the results shown in Figs. 3a and b in the main text. However, as a side effect of the smaller $D_h$, the cell with cathode #1 is expected to experience a larger pressure drop ($\Delta P$) than the case for the cell with cathode #2. For the same $G$, a larger $\Delta P$ results in a larger pumping work ($G\Delta P$ [W]) necessary to force the liquid through the cell. This point is discussed in Section 3.5 of the main text.

## 3. Assessment of temperature non-uniformity on the cathode surface

Here we assess the extent of the temperature non-uniformity on the liquid-contacting plane of the cathode based on a theoretical analysis and computational simulation.

First, the thermal conductivity ($k$) of pure nickel (the cathode material; see Section 5 of the main text) at 400 K is ca. 80 W/m·K and the heat transfer coefficient ($h$) on the cathode surface in this report was 100–600 W/m²·K (see Fig. 3e of the main text). The distance between the hottest and coldest points in a solid under consideration is usually chosen as the representative length ($L$) to calculate the Biot number ($Bi = hL/k$).[S3] Thus, in the present study, $L$ is half of the length of the cathode (i.e., 26.9/2 ≅ 13.5 mm). Using these values, $Bi$ corresponding to $h$ = 100–600 W/m²·K was calculated to be 0.02–0.1, which is smaller than unity. This physically indicates that the *conductive* thermal resistance in the cathode is minor compared to the *convective* thermal resistance at the liquid–solid interface.

To carry out more specific analysis, we introduce thermal resistances $R_{cathode,parallel}$, $R_{cathode,normal}$, and $R_{convection}$ as defined in Fig. S2, which are the conductive thermal resistances in the cathode in the directions parallel to the flow and normal to the flow and the convective thermal resistance between the cathode and liquid, respectively. We also introduce the cross-sectional areas $A_{parallel}$ and $A_{normal}$ as defined in Fig. S2. For the aforementioned conditions, these thermal resistances are calculated to be $R_{cathode,parallel}$ = 2.1 K/W, $R_{cathode,normal}$ = 0.08 K/W, and $R_{convection}$ = ca. 3–17 K/W. Therefore, the conductive thermal resistance inside the cathode is smaller than the convective thermal resistance at the liquid–solid interface. This again indicates that the temperature distribution in the cathode is smaller than the temperature difference between the liquid and cathode, although for the condition of the highest $G$ ($G$ ≅ 0.5 mL/s corresponding to $h$ ≈ 600 W/m²·K), $R_{cathode,parallel}$ becomes closer to $R_{convection}$.

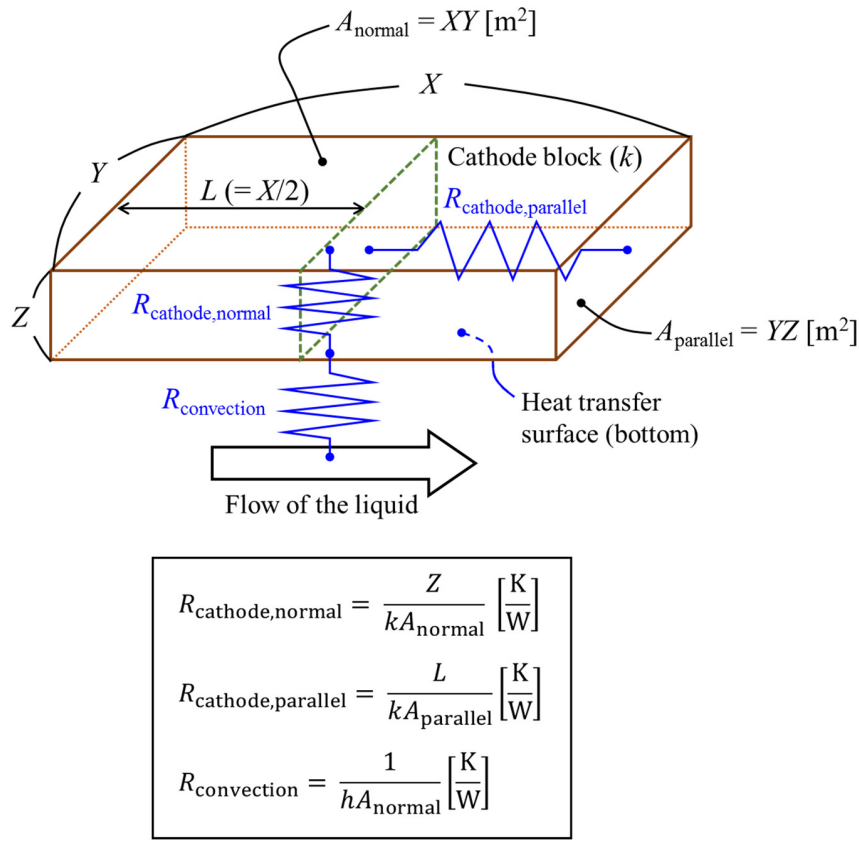

$$R_{\text{cathode,normal}} = \frac{Z}{kA_{\text{normal}}} \left[\frac{K}{W}\right]$$

$$R_{\text{cathode,parallel}} = \frac{L}{kA_{\text{parallel}}} \left[\frac{K}{W}\right]$$

$$R_{\text{convection}} = \frac{1}{hA_{\text{normal}}} \left[\frac{K}{W}\right]$$

**Fig. S2** Schematic description of the model used to analyze the thermal resistances relevant to the heat transfer in the cathode and that between the cathode and liquid. $X$, $Y$, and $Z$ represent the dimensions of the cathode, where $X$ is parallel to the direction of the flow. Definitions of the variables are also given.

Next, we carry out a simulation for the cell with cathode #1 at $G = 0.5$ mL/s and $T_{\text{cathode}} = 170$ °C, corresponding to the conditions that gave the maximum temperature non-uniformity in cathode #1 in the above analysis. Figure S3 presents the simulated temperature distribution on the liquid-contacting plane of the cathode. This figure indicates that the temperature variation over most of the area (corresponding to the colors between red and light green) falls within ca. 6 K, or ca. ±3 K around the temperature at the thermocouple position (which is in the orange region in the color contour plot). This non-uniformity was sufficiently smaller than the temperature difference between the cathode and anode (ca. 130 K for these conditions, see the main text for details).

Therefore, we assume that the temperature of the cathode can be represented by the temperature measured by the thermocouple embedded in the cathode throughout this report.

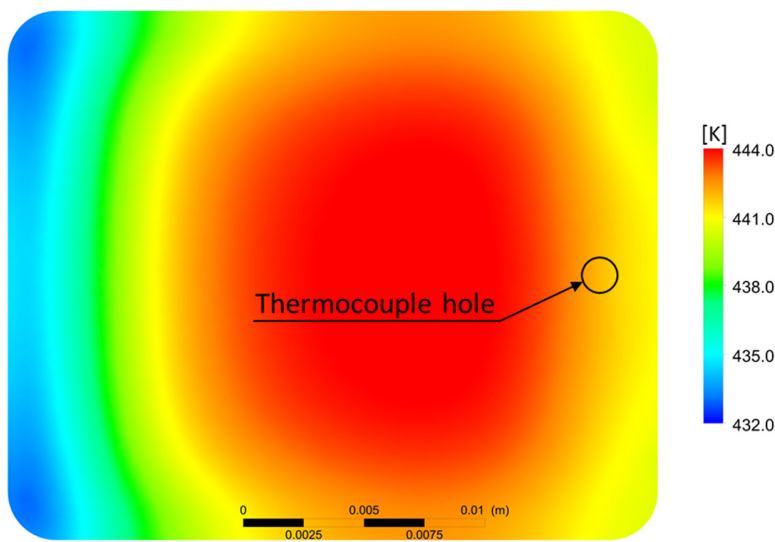

**Fig. S3** Simulated temperature distribution on the liquid-contacting plane of cathode #1 for $T_{\text{cathode}}$ = 170 °C and $G$ = 0.5 mL/s. In this graphic, the liquid flow was from left to right. The position of the hole for the thermocouple used to measure $T_{\text{cathode}}$ is indicated by a circle.

## 4. Derivation of the small-signal mass transfer resistance $R_{\text{mt}}$

Full details of a semi-empirical model for steady-state mass transfer have been given in Section 1.4 of a textbook by Bard and Faulkner.[S4] In this model, a redox reaction on the electrode surface O + $n$e ↔ R (O: oxidized form; R: reduced form; $n$: the number of electrons involved) is considered for the situation where a steady-state flow of an electrolyte parallel to the electrode plane (which may be caused by stirring or forced convection) has formed a Nernst diffusion layer on the electrode,[S4] as depicted in Fig. S4. This figure depicts an example situation of a reaction O + $n$e → R where the electrode is a cathode. The concentration of the redox species (O in this case) outside the diffusion layer of thickness $\delta_O$ is maintained at the bulk concentration ($C_O^*$) because of the steady flow of the electrolyte, whereas the concentration on the electrode surface ($C_{O(x=0)}$) is lower than $C_O^*$ because of the steady consumption of O by the Faradaic reaction occurring at the electrode. $C_O$ varies with the distance ($x$) from the electrode in the diffusion layer.

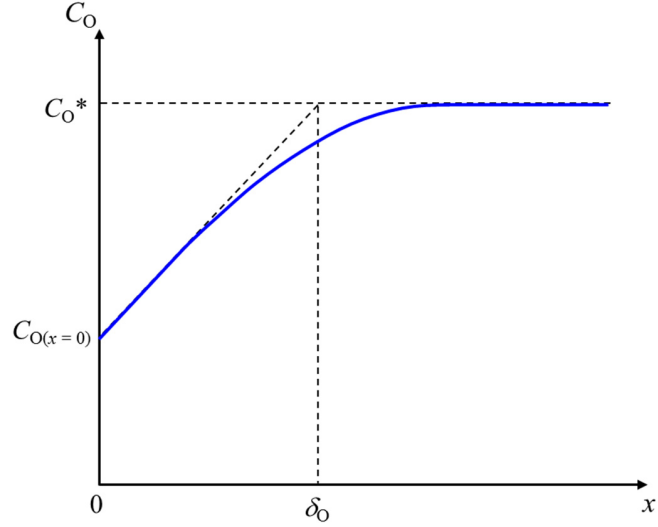

**Fig. S4** Concentration profile for species O near the electrode. The electrode surface is $x = 0$. The solid curve represents the concentration profile and the dashed lines represent the diffusion layer approximation. $\delta_O$ and $C_{O(x=0)}$ are the diffusion layer thickness and concentration of O on the electrode, respectively. This figure was drawn based on Figure 1.4.1 of ref. S4.

Because the details of this mass transfer model have been described elsewhere,[S4] here we only briefly introduce this model to derive a solution (eqn (5) in the main text). First, the current–potential ($I$-$E$) relationship for the case where both O and R are present initially is given by[S4]

$$E = E^{0\prime} - \frac{RT}{nF}\ln\left(\frac{m_O}{m_R}\right) + \frac{RT}{nF}\ln\left(\frac{I_{\mathrm{lim,c}}-I}{I-I_{\mathrm{lim,a}}}\right), \tag{S1}$$

where $E^{0\prime}$ is the formal potential, $R$ is the gas constant, $T$ is the absolute temperature, $F$ is the Faraday constant, $I_{\mathrm{lim,c}}$ is the cathodic limiting current, and $I_{\mathrm{lim,a}}$ is the anodic limiting current. Here, the variables $m_X$ (X: O or R) are defined by

$$m_X \equiv \frac{D_X}{\delta_X}, \tag{S2}$$

where $D_X$ is the diffusion coefficient of species X. In eqn (S1), the signs of $I_{\text{lim,c}}$ and $I_{\text{lim,a}}$ are defined to be opposite to each other.[S4] From eqn (S1), the equilibrium potential $E_{\text{eq}}$, which is the potential at $I = 0$, can be expressed by

$$E_{\text{eq}} = E^{0\prime} - \frac{RT}{nF}\ln\left(\frac{m_\text{O}}{m_\text{R}}\right) + \frac{RT}{nF}\ln\left(-\frac{I_{\text{lim,c}}}{I_{\text{lim,a}}}\right), \tag{S3}$$

Then, the mass transfer overpotential $\eta_{\text{mt}}$ can be expressed as[S4]

$$\eta_{\text{mt}} = E - E_{\text{eq}} = \frac{RT}{nF}\ln\left(\frac{I_{\text{lim,c}}-I}{I-I_{\text{lim,a}}}\right) - \frac{RT}{nF}\ln\left(-\frac{I_{\text{lim,c}}}{I_{\text{lim,a}}}\right). \tag{S4}$$

As a result of the subtraction in eqn (S4), the parameters $m_X$ (and thus $D_X$ and $\delta_X$) are eliminated from this equation. For the conditions of small deviations of $E$ from $E_{\text{eq}}$ (i.e., $E \cong E_{\text{eq}}$), the *"small-signal" mass transfer resistance $R_{\text{mt}}$*[S4] can be defined as

$$R_{\text{mt}} \equiv \left(\frac{\partial \eta_{\text{mt}}}{\partial I}\right)_{I=0}. \tag{S5}$$

By substituting eqn (S4) into eqn (S5), $R_{\text{mt}}$ is obtained as follows

$$R_{\text{mt}} = \frac{RT}{nF}\left(\frac{I_{\text{lim,a}}-I_{\text{lim,c}}}{I_{\text{lim,a}}I_{\text{lim,c}}}\right). \tag{S6}$$

Equation (S6) is the representation for the situation of Fig. S4 and thus the terms "$I_{\text{lim,c}}$" and "$I_{\text{lim,a}}$" have been defined for one electrode. However, in the present forced-flow thermocell, because the mass transfer resistance contained contributions from both the cathode and anode, an overall $R_{\text{mt}}$ has to be determined. To consider this point, the pair of terms representing the two limiting currents above ($I_{\text{lim,c}}$ and $I_{\text{lim,a}}$) are replaced by a new pair of terms, $I_{\text{lim,forward}}$ and $I_{\text{lim,backward}}$. Here, "$I_{\text{lim,forward}}$" is the limiting current found from Figs. 4d and 5a in the main text, where our "cathode" and "anode" literally work according to their names. Conversely, "$I_{\text{lim,backword}}$" is the limiting current that would be measured when our "cathode" and "anode" work as an anode and cathode, respectively; i.e., when the current flows inversely to the intended design by applying an inverse potential between the electrodes. Using the SMU (see Section 5 of the main text), such an inverse measurement was possible. Figure S5 shows the *I-V* curves acquired for the normal and inverse scans obtained using cathode #2 with $T_{\text{cathode}}$ = 170 °C and $G$ = 0.5 mL/s. Figure S5 shows that $I_{\text{lim,backward}}$ was equal to $-I_{\text{lim,forward}}$ in the present cell, and therefore the following relation can be used for the present situation

$$I_{\text{lim,forward}} \cong -I_{\text{lim,backward}} \equiv I_{\text{lim}} \, (> 0). \tag{S7}$$

By substituting eqn (S7) into eqn (S6), $R_{\text{mt}}$ can be expressed as

$$R_{\text{mt}} \cong \frac{RT}{nF} \frac{2}{I_{\text{lim}}}, \tag{S8}$$

which is eqn (5) in the main text.

Next, the reason for the factor "2" in eqn (S8) is briefly mentioned. As stated in the main text, $R_{\text{mt}}$ in this work is the *overall* mass transfer resistance containing contributions from diffusion layers on both the cathode and anode. However, the factor "2" in eqn (S8) did not originate from this aspect. The essential feature of eqn (S1) is that the magnitude of $R_{mt}$ near $E \cong E_{eq}$ (or $I \cong 0$) contains contributions from both the limiting current for the forward reaction and that for the reverse reaction when the reduced and oxidized species coexist in a liquid electrolyte.[S4] Therefore, the factor "2" in eqn (S8) originates from the relationship in eqn (S7).

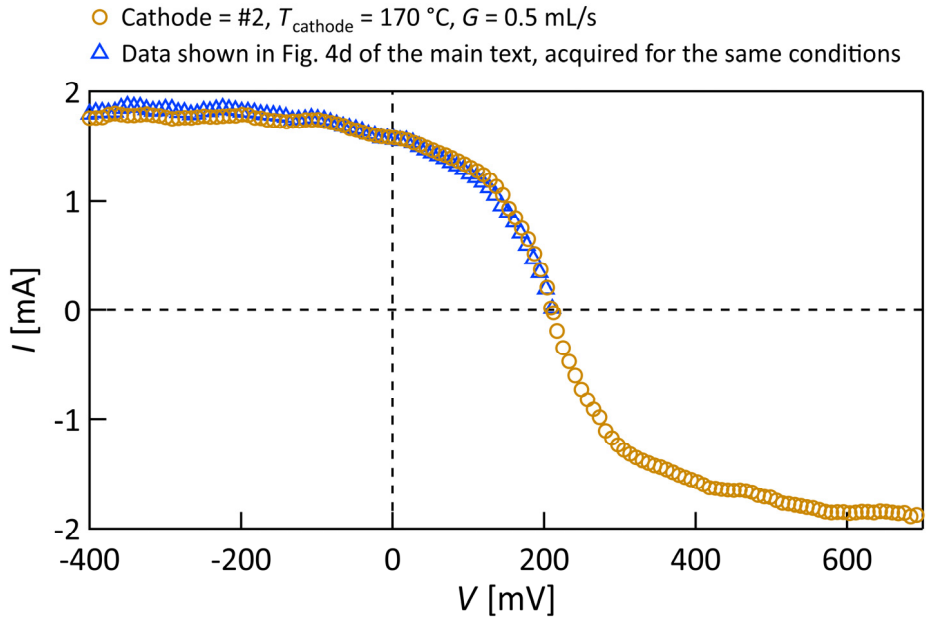

**Fig. S5** *I-V* curves acquired by scans in both the normal and inverse directions using cathode #2 with $T_{\text{cathode}}$ = 170 °C and $G$ = 0.5 mL/s (circles). These normal and inverse scans were carried out on the same day. As a comparison, the *I-V* curve scanned only for the normal direction for the same experimental conditions is also shown (triangles), which is also presented in Fig. 4d of the main text. The acquisition of the former was carried out a few months later than that of the latter, demonstrating the high reproducibility of the experimental results.

As briefly mentioned in the main text, Abraham et al.[S5] previously used the same theoretical framework in their report to obtain $R_{mt}$ but they followed a different route from ours presented above. In their route, $I_{lim}$ was calculated using

$$|I_{lim}| = \frac{nFADC}{\delta}, \tag{S9}$$

which is equation (1.4.17) in the textbook authored by Bard and Faulkner.[S4] Specifically, Abraham et al.[S5] used the value of $D$ separately obtained from their experiments while treating $\delta$ as an adjustable parameter in eqn (S9). In our case, $I_{lim}$ was obtained experimentally and thus it was not necessary to use $D$ or $\delta$ in our route described above. We believe that both routes are equally important because one can choose to use whichever is the better route depending on the available experimental data.

Back to our analysis, eqn (S4) can be rewritten using eqn (S7) as

$$E = E_{eq} + \frac{RT}{nF} \ln\left(\frac{I_{lim} - I}{I_{lim} + I}\right), \tag{S10}$$

which describes the relationship between the voltage and current as a function of $I_{lim}$. An example of the fit to the experimental $I$-$V$ curves in Fig. 5a of the main text with eqn (S10) is shown in Fig. S6. Further discussion and analyses of eqn (S10) are beyond the scope of the present report.

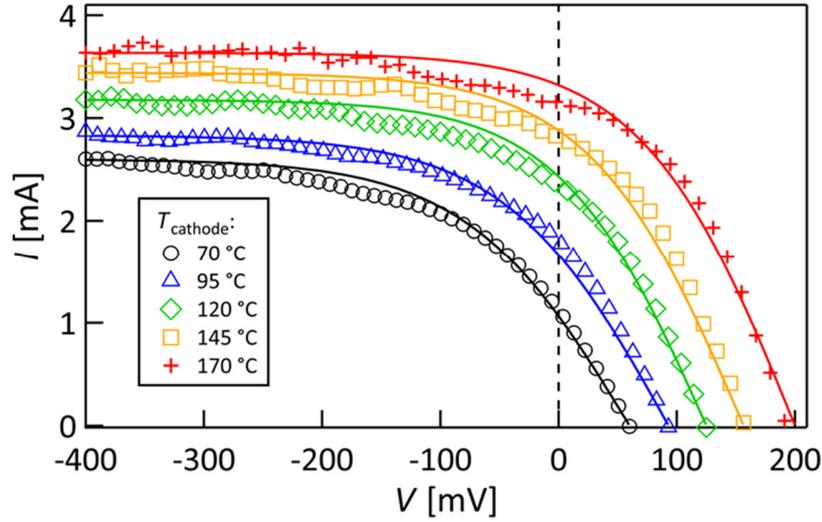

**Fig. S6** Example fits of eqn (S10) to the *I-V* curves shown in Fig. 5a of the main text.

## 5. Temperature dependence of the viscosity of the working liquid

The temperature dependence of the viscosity of the working liquid was measured using a cone-plate rheometer equipped with a thermoelectrically controlled temperature stage (R/S Plus, Brookfield). The measurements were carried out under dry nitrogen gas to avoid the effect of moisture on the viscosity of the working liquid. Figure S7 shows the results along with the theoretical fit by the Vogel–Fulcher–Tammann (VFT) equation[S6]

$$\eta(T) = A \exp\left(\frac{B}{T-C}\right), \tag{S11}$$

where *A*, *B*, and *C* are the parameters. Our fit yielded $A = 7.2 \times 10^{-4}$ Pa·s, $B = 496.4$ K, and $C = 179.7$ K. These parameters were used to calculate the viscosity of the working liquid in our simulations. It should be noted that the viscosity obtained in Fig. S7 was higher than that of the neat ionic liquid, as shown later in Fig. S12.

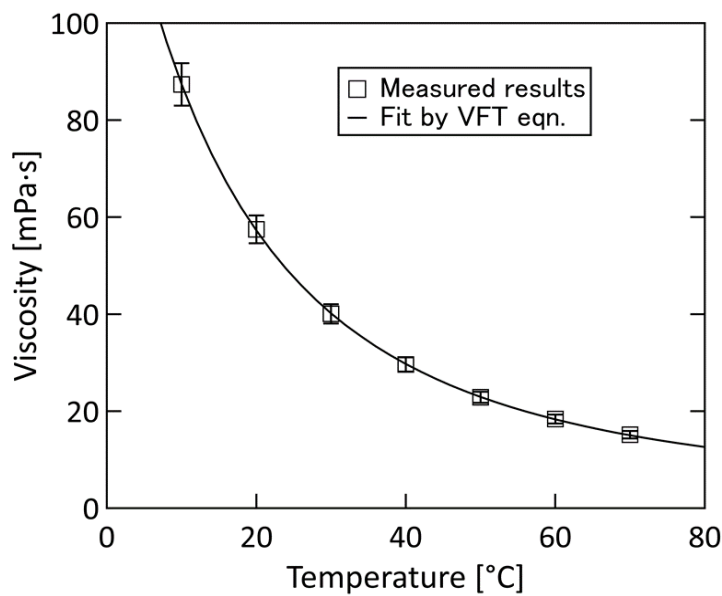

**Fig. S7** Temperature dependence of the viscosity of the working liquid used in this report measured under dry nitrogen gas. The error bars represent ±5% uncertainty in the measurements. The black curve is a theoretical fit by the VFT equation.

## 6. Arrhenius plots of the working liquid viscosity and $R_{ct}^{-1}$

Figures S8a and b show two Arrhenius plots to support the discussion in the main text. Figure S8a is the Arrhenius plot for the viscosity of the working liquid, which was generated from our data presented in Fig. S7. The slope represents the activation energy of the viscosity, $E_{a(\eta)}$. Figure S8b is the Arrhenius plot for the inverse of $R_{ct}$ (data shown in Fig. 5d of the main text), where the slope represents the apparent activation energy.[S7] For the temperature for the latter, the average temperature of $T_{cathode}$ and $T_{anode,ave}$ was used, where $T_{anode,ave}$ was obtained from our simulations. In agreement with the results of Tachikawa et al.,[S7] who investigated the electrode kinetics of several ion complexes in an ionic liquid similar to [C$_2$mim][NTf$_2$] used in the present report, these activation energies are similar to each other.

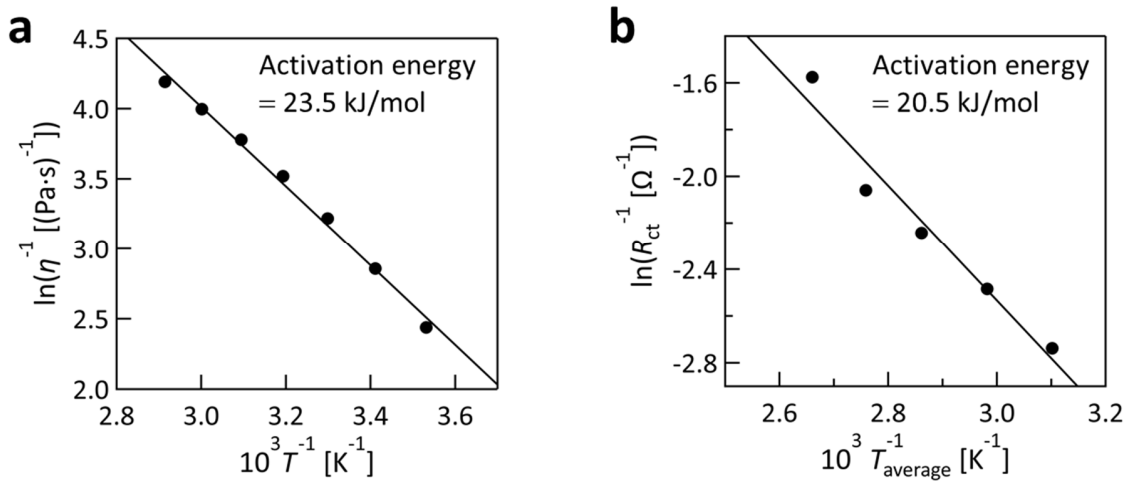

**Fig. S8** Arrhenius plots of (a) the viscosity of the working liquid and (b) the inverse of $R_{ct}$ obtained from the temperature-dependent experiments shown in Fig. 5 of the main text.

## 7. Dependence of $T_{anode,ave}$ on $G$

To support the explanation given for the increase of $V_{OC}$ with increasing $G$ found in Fig. 6a of the main text, corresponding simulations were carried out for the condition of $T_{cathode}$ = 170 °C. Figure S9 shows the simulated relation between $T_{anode,ave}$ and $G$. As expected, $T_{anode,ave}$ decreased with increasing $G$, supporting our explanation for the observed increase of $V_{OC}$ with $G$ in the main text.

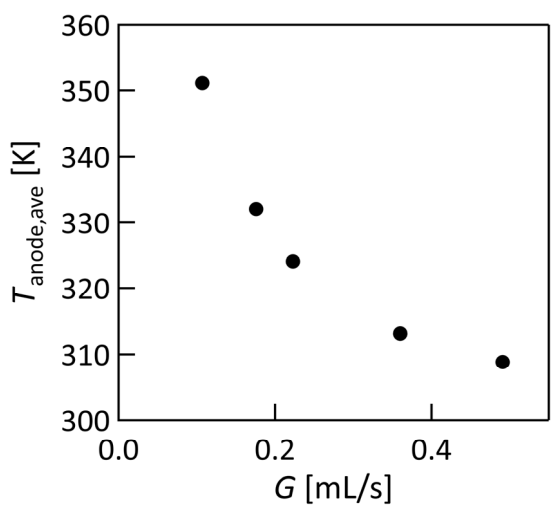

**Fig. S9** Simulated $T_{anode,ave}$ vs. $G$ for the experimental conditions in Fig. 6 in the main text.

## 8. Randles plots generated from the results of AC impedance measurements

Using the results of AC impedance measurements, plots of impedance vs. $\omega^{-1/2}$ ($\omega$: AC frequency), often called Randles plots, can be generated. Randles plots are useful for analyzing the electrochemical processes.[S4] As described in Section 10.3 of ref. S4 as well as Section 3.4 of the main text, the slope of a Randles plot is related to the diffusion coefficient of the redox species in the solvent (eqn (6) in the main text). Figure S10 shows the Randles plots for different $G$ generated from the data presented in Fig. 6d of the main text. In all the panels in Fig. S10, the plots for both the real ($Z_{Re}$) and imaginary ($Z_{im}$) components are linear and have a common slope, which indicates that the present system can be described by the equivalent circuit model.[S4]

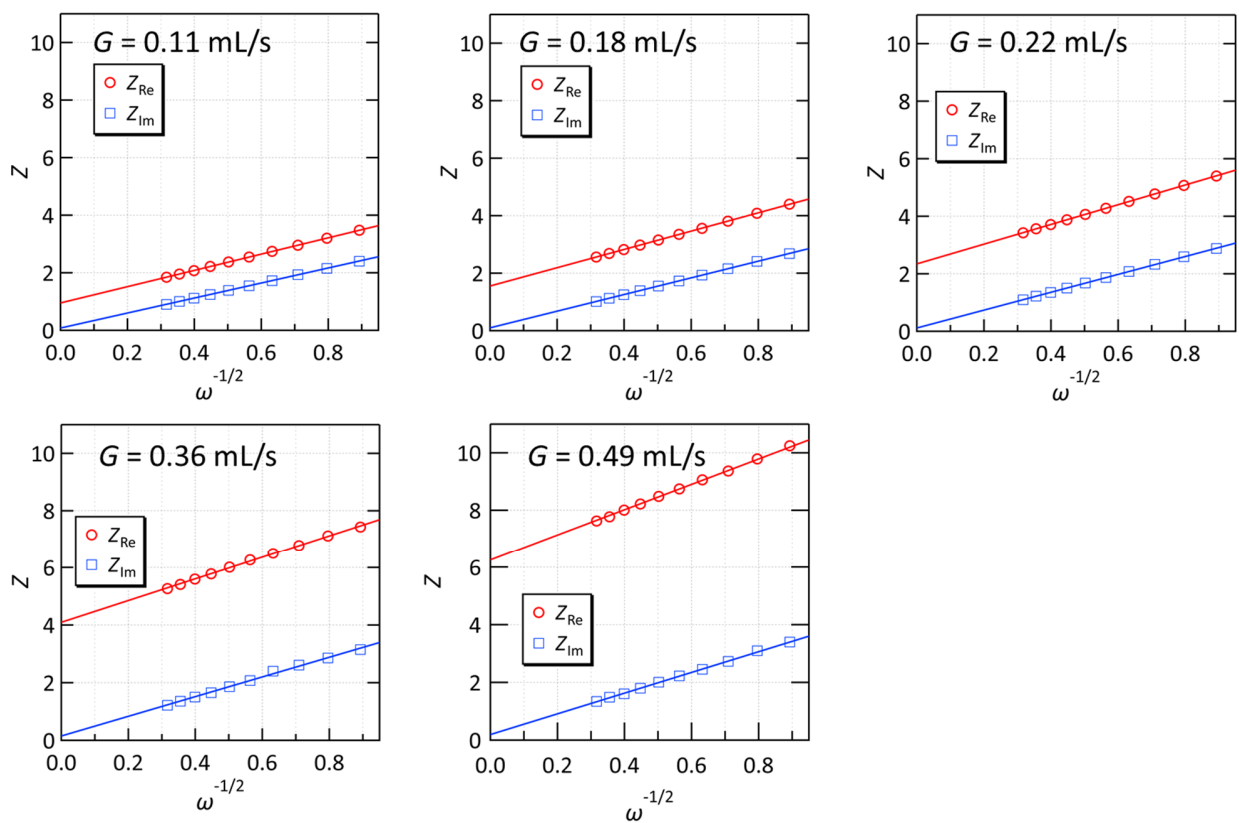

**Fig. S10** Randles plots generated from the results of the AC impedance measurements shown in Fig. 6d of the main text.

## 9. UV-vis optical absorption spectrum of [C$_2$mim][NTf$_2$]

Figure S11 shows the optical absorption spectrum of the ionic liquid used in this report, [C$_2$mim][NTf$_2$] supplied from IoLiTec, Germany (see also Section 5 in the main text). The observed weak optical absorption in the UV region and the absence of absorption in the visible region supported the high purity of [C$_2$mim][NTf$_2$] certified by the supplier (> 99.5%). All optical absorption measurements in this report were conducted using a UV−vis−near-infrared spectrophotometer (UV-3600, Shimadzu) with a 1 mm-thick quartz cell.

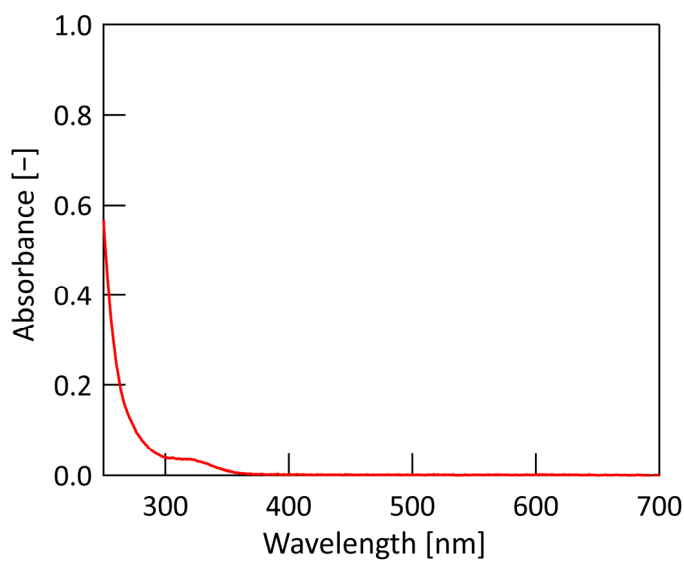

**Fig. S11** Optical absorption spectrum of [C$_2$mim][NTf$_2$] used in this report. The optical path length was 1 mm.

# 10. Temperature dependence of the viscosity of [C$_2$mim][NTf$_2$]

The temperature dependent viscosity of the ionic liquid [C$_2$mim][NTf$_2$] used in this report was also measured with the same equipment used to obtain Fig. S7. The measurements were carried out under dry nitrogen gas to avoid the effect of moisture on the viscosity of [C$_2$mim][NTf$_2$]. Prior to the measurements, the ionic liquid was dried under vacuum at 120 °C. Figure S12 compares the obtained results (circles) with previously reported results (squares and triangles).[S8,S9] The black curve is a theoretical fit of the VFT equation (eqn (S11)) to our results.

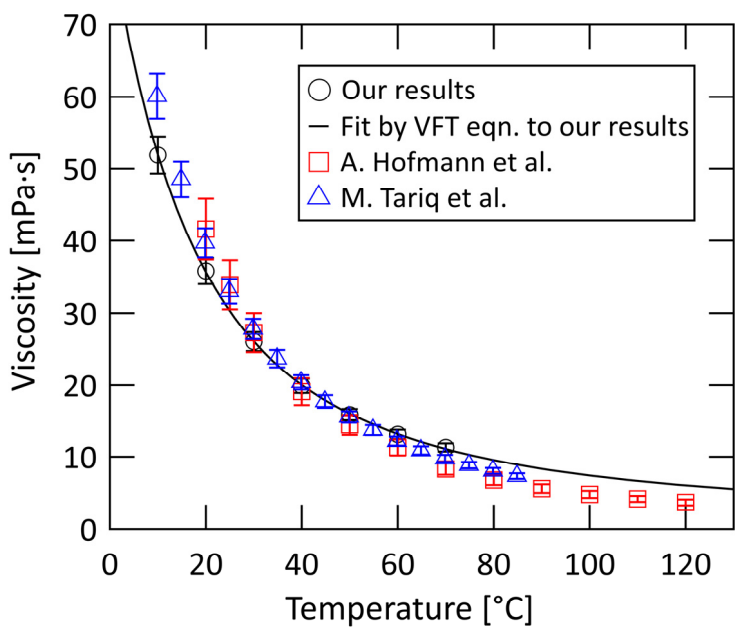

**Fig. S12** Temperature dependence of the viscosity of [C$_2$mim][NTf$_2$] used in this report (circles) measured under dry nitrogen. The squares and triangles are the data previously reported in Ref. S8 and S9, respectively. The error bars for our results represent ±5 % uncertainty in the measurements, whereas the error bars for the previous results are the uncertainties given in these reports. The black curve is the fit of the VFT equation to our results.

## 11. UV-vis optical absorption spectrum of the working liquid

Figure S13 shows the optical absorption spectrum of the working liquid (0.06 M solution of $Co^{II/III}(bpy)_3(NTf_2)_{2/3}$ in $[C_2mim][NTf_2]$) prepared using $Co^{II/III}(bpy)_3(NTf_2)_{2/3}$ supplied by Nippon Kayaku, Japan (black curve). To check the quality of the redox couple, we also purchased $Co^{II/III}(bpy)_3(NTf_2)_{2/3}$ from a different supplier (Dyenamo, Sweden) and compared the optical absorption spectrum of the solution (pink dashed curve) with that used in this report. Figure S13 shows that the agreement between these two spectra was excellent in terms of both the spectral shape and quantitative absorbance values, supporting the high purity of the redox couple used in this report.

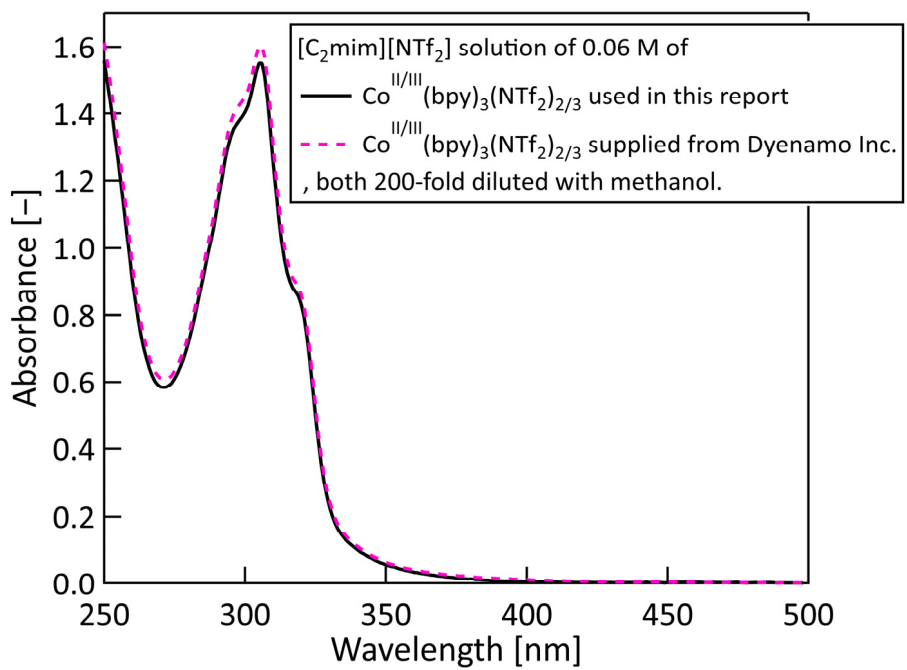

**Fig. S13** Optical absorption spectrum of the working liquid (0.06 M solution of $Co^{II/III}(bpy)_3(NTf_2)_{2/3}$ in $[C_2mim][NTf_2]$) used in this report (black curve) compared with that of a working liquid prepared using $Co^{II/III}(bpy)_3(NTf_2)_{2/3}$ purchased from a different supplier (Dyenamo, Sweden; pink dashed curve). Both spectra were measured after 200-fold dilution with methanol. The optical path length was 1 mm.

## 12. Details of the SMU settings

We found that proper choice of the measurement parameters in the source measure unit (SMU) was important for acquiring sound *I-V* curves. In particular, we found that improper choice of the parameters can cause artifacts, which are falsely better *I-V* curves and higher generated powers. Therefore, we present here the settings of the SMU used in this report. The SMU used in this study was a Model 2450 from Keithley. This SMU has two main parameters, "filter count (FC)" and "number of power line cycles (NPLCs)" that set the signal averaging time. As these values become larger, the time to complete an acquisition of one *I-V* curve becomes longer.

Figure S14 shows the dependences of $P_{max}$ acquired in preliminary experiments carried out at $G = 0.12$ and $0.45$ mL/s on the values of NPLCs and FC. The choice of small FC and NPLCs resulted in a falsely larger $P_{max}$, which was presumably caused by the transient current due to insufficient achievement of the steady state of the electrode reactions. Throughout this report, we used NPLCs = 10 and FC = 50, with which the steady state of the electrode reactions was sufficiently achieved and aforementioned artifacts could be avoided.

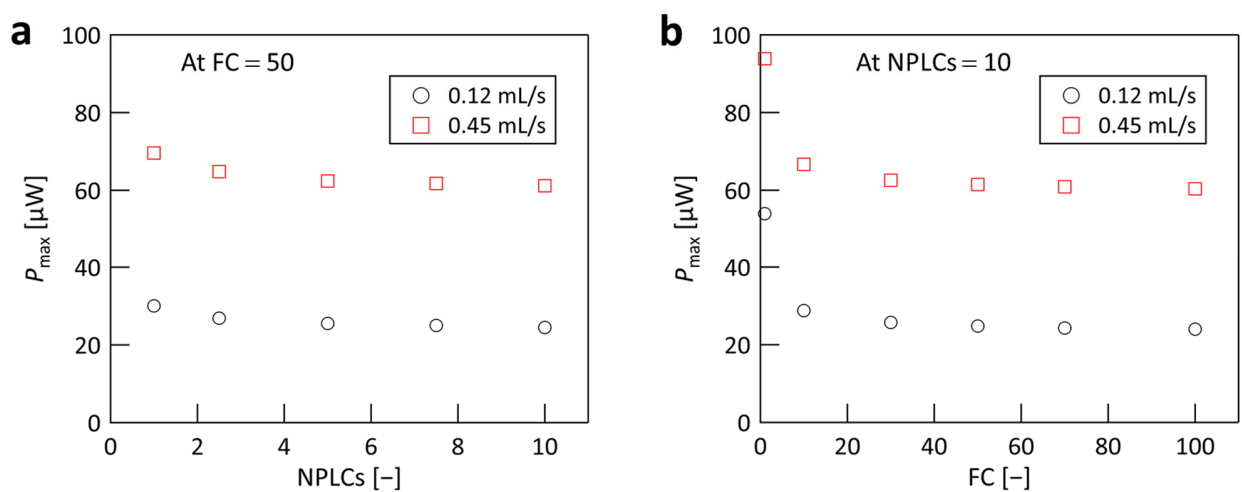

**Fig. S14** Dependence of $P_{max}$ obtained from the *I-V* curves acquired using the SMU on the parameters of (a) NPLC and (b) FC measured under FC = 50 and NPLCs = 10, respectively, at flow rates of 0.12 (circles) and 0.45 mL/s (squares).